\documentclass[aps,nofootinbib,twocolumn,superscriptaddress]{revtex4-1} 
\usepackage{graphicx}

\usepackage[english]{babel}
\usepackage{amssymb,amsfonts,psfrag,amsmath,bm,bbm,indentfirst,subfigure,bbold}
\usepackage{color}
\usepackage{ulem}
\usepackage{natbib}

\RequirePackage[
   hyperindex,colorlinks,bookmarksnumbered,
   plainpages=true,pdfstartview=FitH]{hyperref}
\hypersetup{linkcolor=blue,urlcolor=blue,citecolor=blue}
\usepackage{hyperref}
\usepackage{breakurl}

\definecolor{darkgreen}{rgb}{0,0.5,0}
\definecolor{purple}{rgb}{0.35,0,0.35}
\definecolor{orange}{rgb}{1,0.5,0}
\definecolor{darkred}{rgb}{.7,0,0}
\definecolor{darkblue}{rgb}{0.1,0.1,.6}
\definecolor{grey}{rgb}{.6,.6,.6}
\definecolor{dimgreen}{rgb}{0.2,0.6,0.1}
\definecolor{RoyalBlue}{cmyk}{0.94,0.539,0,0}
\definecolor{DGLorange}{cmyk}{.22,1,1,.2}
  
\newcommand{\Chi}{{\rm X}}
\newcommand{\Tr}{{\operatorname{Tr}}}

\newcommand{\jvd}[1]{{\color{black}{#1}}}

\newcommand{\changed}[1]{{\color{black}{#1}}}

\newcommand{\Imp}{\operatorname{Im}}
\newcommand{\Rep}{\operatorname{Re}}

\renewcommand{\emph}[1]{\textit{#1}}
\begin{document}
\allowdisplaybreaks

\title{Keldysh Functional Renormalization Group Treatment of Finite-Ranged Interactions in Quantum Point Contacts}

\author{Lukas Weidinger}
\author{Jan von Delft}

\affiliation{Arnold Sommerfeld Center for Theoretical Physics and Center for
NanoScience,
Ludwig-Maximilians-Universit\"at
M\"unchen, Theresienstrasse 37, D-80333 M\"unchen, Germany}

\date{\today}

\begin{abstract}
We combine two recently established methods, the extended Coupled-Ladder Approximation (eCLA) [Phys. Rev. B \textbf{95}, 035122 (2017)] and a dynamic Keldysh functional Renormalization Group (fRG) approach for inhomogeneous systems [Phys. Rev. Lett. \textbf{119}, 196401 (2017)] to tackle the problem of finite-ranged interactions in quantum point contacts (QPCs) at finite temperature.
Working in the Keldysh formalism, we develop an eCLA framework, proceeding from a static to a fully dynamic description. 
Finally, we apply our new Keldysh eCLA method to a QPC model with finite-ranged interactions and show evidence that an interaction range comparable to the length of the QPC might be an essential ingredient for the development of a pronounced 0.7-shoulder in the linear conductance.  
We also discuss problems arising from a violation of a Ward identity in second-order fRG.
\end{abstract}

\maketitle

\section{Introduction}
In a previous work \cite{Weidinger2017}, we have devised an extended Coupled-Ladder Approximation (eCLA), an approximation scheme within the second-order truncated functional Renormalization Group (fRG) approach. The eCLA is capable of a controlled incorporation of the spatial extent of the one-particle irreducible two-particle vertex (hereafter simply called ''vertex``) into a channel-decomposed \cite{Karrasch2008,Jakobs2010,Bauer2013} fRG flow. 
Using a static Matsubara implementation, we showed that this scheme improves the convergence of the fRG flow by increasing the feedback between the separate channels of the vertex flow. 
Furthermore, by design, this scheme includes a correct treatment of finite-ranged interactions up to second order in the interaction. 
Applying the eCLA scheme to a quantum point contact (QPC), we observed that with an increasing interaction range, the effective QPC barrier flattens and additional features in the linear conductance (herafter simply called ''conductance``) arise, caused by corresponding Friedel oscillations.

The eCLA has recently also been used in \cite{Markhof2018} to study phase transitions in an one-dimensional spinless tight-binding chain with nearest and next nearest neigbor interaction. 
Furthermore, in \cite{Sbierski2017} a set of second order flow-equations was derived for a one-dimensional system of spinless fermions, which can be obtained as a special case of the spin-1/2 eCLA equations. 
 
In this paper, we build on our previous QPC studies, now focusing on the following question: how does the temperature dependence of the QPC conductance change when the interaction range is increased from $0$ up to the scale of the characteristic QPC length?  
In this regime, our previous zero-temperature static Matsubara approach indicated only a slight broadening of the conductance step. 
However, it is very interesting to study the behavior in this regime at finite temperature, \changed{since -- contrary to experimental findings, see e.g.\ \cite{Thomas1996,Bauer2013,Iqbal2013} -- an earlier study \cite{Schimmel2017}, utilizing only onsite interactions, found no pronounced 0.7-shoulder in the conductance. }
In order to be able to treat finite temperatures, we here present an implementation of the eCLA in a dynamic Keldysh setup, as devised in \cite{Jakobs2009,Jakobs2010a} and extended and successfully applied to QPCs with short-range interactions in \cite{Schimmel2017}.
Since a full treatment of both the spatial as well as the frequency structure of the vertex is numerically not possible, we introduce an additional approximation scheme that allows us to take the extended spatial structure of the vertex for successively more frequencies into account.
Although the numerical costs did not permit us to reach full convergence w.r.t.\ the used frequency range, the qualitative behavior at large ranges remained stable.   
Furthermore, we analytically argue that we are indeed able to capture the most important vertex contributions to the conductance within the covered frequency range. 

Finally, we apply this new method to a QPC at finite temperature and show evidence that a finite interaction range on the scale of the length of the QPC likely is an essential factor for the development of a pronounced 0.7-shoulder in the conductance (see Fig.~\ref{fig_dyn_temp_dep} below).  

We also discuss problems arising from a violation of a Ward identity in second-order fRG. 
We suggest a simple correction factor for ameloriating these problems, but conclude that a truly reliable cure will require going beyond second-order fRG.

This paper is structured as follows. 
Sec.~\ref{Model} defines the model used to describe a QPC.
Sec.~\ref{Method} describes methodological details, in particular regarding our parametrization of the vertex. (Problems arising from a Ward identity violation are addressed in Sec.~\ref{conductance_computation}, see Fig.~\ref{fig_two_pos} below). 
Sec.~\ref{Results} presents our results for the temperature dependence of the QPC conductance and Sec.~\ref{Conclusion} our conclusions.
Three appendices deal with further technical details, such as vertex symmetries (App.~\ref{app_sym}), the importance of a dynamic treatment of vertex feedback (App.~\ref{importance_feedback}), and the consequences of violating Ward identities (App.~\ref{vio_ward_identities}).

\section{Model}
\label{Model}
We consider a Hamiltonian consisting of a one-dimensional tight-binding chain with finite-ranged interactions:
\begin{align}
\label{hamiltonian}
H&=-\sum_{i \sigma} \tau_i [ c^\dagger_{i\sigma} c_{i+1\sigma} + h.c] + \sum_{i \sigma} \sigma \frac{B}{2} n_{i \sigma} \nonumber \\
  &+\tfrac{1}{2} \sum_{i j \sigma \sigma'} U_{ij} (1 - \delta_{ij} \delta_{\sigma \sigma'}) n_{i \sigma} n_{j \sigma'}, 
\end{align} 
where $c_{i\sigma}$ annihilates an electron at site $i \in \mathbb{Z}$ with spin $\sigma$ and $n_{i \sigma} = c^\dagger_{i \sigma} c_{i \sigma}$ is the number operator.
Instead of a quadratic onsite potential as used in \cite{Weidinger2017}, we use a quadratic modulation in the hopping, $\tau_i = \tau - \Delta \tau_i$, to model the QPC barrier. 
This approach was also used in \cite{Schimmel2017}. It causes a constriction of the tight-binding band, leading to a density of states which, close to the lower band edge, is equivalent to the one generated by a quadratic onsite potential. 
Moreover, at the upper band edge this method avoids the formation of sharp bound states which are difficult to treat numerically and lead to problems with e.g. the normalization of the density of states.  

\begin{figure}
   \includegraphics[scale=0.7]{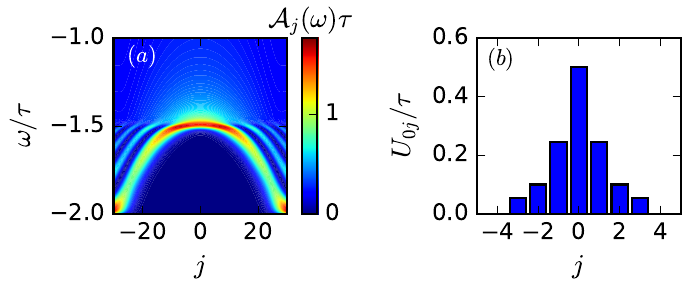} 
   \caption{\small (a) Colorplot of the non-interacting LDOS $\mathcal{A}_j(\omega)$ for the chosen QPC model.  (b) Interaction profile $U_{0j}$ in the center of the QPC as function of site $j$.}
\label{fig_model}
\end{figure}

The hopping modulation and the interactions are both taken to be finite only within a central region with $2N+1$ sites, i.e. $U(i,j)=0$, if $i$ or $j$ $\not\in [-N,N]$ and $\Delta \tau_i=0$ if $i \not\in [-N,N-1]$. 
Note that the central region contains one hopping element less than onsite terms. 
Within this region the hopping and interaction takes the form  
\begin{align}
\Delta \tau_{i } =& \frac{1}{2} V_g e^{-x_i^2/(1-x_i^2)}, \ x_i = \frac{2i+1}{2N}, \label{hopping_variation} \\
U_{ij}       =& \Big[\delta_{ij} U_0 + (1-\delta_{ij}) U_1 \frac{e^{-|i-j|/\chi}}{|i-j|} \Big] f(i,j),  \label{Yukawa} 
\end{align}
where $i \in [-N,N-1]$ for $\Delta \tau_i$ and $i,j \in [-N,N]$ for $U_{ij}$. 
\changed{The hopping variation $\Delta \tau_j$ is characterized by $V_g$, the effective barrier height in the center of the QPC, as well as an exponential factor $\exp[-x_i^2/(1-x_i^2)]$ governing the form of the barrier: In the QPC center a quadratic barrier top dominates, while in the QPC flanks the barrier goes smoothly to zero.  
The interaction consists of an onsite term $\delta_{ij}U_0$ as well as a Yukawa-like offsite term governed by interaction strength $U_1$ and exponential decay rate $\chi$. 
}
%Furthermore, $V_g$ is the effective barrier height in the center of the QPC, $U_0$ denotes the onsite interaction strength and $U_1$ sets the offsite interaction strength, while $\chi$ governs its exponential decay. 
\changed{
We chose the Yukawa-like form of the interaction strength in order to fit two demands: 
(i) The interaction should not only be onsite anymore (as it was in \cite{Schimmel2017}), but also have a finite extent comparable to the characteristic QPC length.   
(ii) It still has to decay quickly enough, i.e.\ not develop an algebraic long-range tail, in order to be numerically treatable at finite temperature within a dynamic Keldysh setup.
The situation of weaker screening, introducing only an algebraic decay in the interaction strength, requires a very large spatial extent of the vertex. 
For this situation, a dynamic treatment within the eCLA approach is therefore not feasible. 
However, for zero temperature, this case can be studied approximately within a static fRG approach that requires considerably lesser numerical resources \cite{Weidinger2017}.}
The function $f(i,j)$ is inserted for numerical purposes and consists of two factors  
\begin{align}
f(i,j) = \exp \Big( -\frac{z(i,j)^6}{1-z(i,j)^2} \Big) \times \theta \Big(L_U - |i-j| \Big),
\label{interaction_dampening_and_cutoff}
\end{align}
with $z(i,j)=\max \Big( \frac{|i|}{N}, \frac{|j|}{N}\Big)$. 
The exponential factor suppresses the interaction at the edges of the central region and thus assures a smooth transition from finite interaction strength to zero interaction in the leads. 
\changed{
Note that instead of the quadratic power that appears in the numerator of the exponential factor in the hopping variation \eqref{hopping_variation}, we used in \eqref{interaction_dampening_and_cutoff} a power of $z(i,j)^6$ in the exponential term.  
This ensures that the interaction strength around the barrier top stays almost constant and only drops off, smoothly, relatively close to the edges of the central region.}
The $\theta$ factor introduces a cutoff in the interaction range, i.e. the interaction is only finite for ranges $|i-j|\leq L_U$.
Since in this work we will focus only on qualitative predictions, we will in fact use only $L_U$ to vary the range of the interaction, while keeping $\chi$ fixed on the scale of the QPC length. 
Concretely, if not specified otherwise, we will use the following parameters throughout:
Spatial discretization $N=30$, i.e. we have a total number of $2N+1=61$ sites;
barrier height $V_g=0.5\tau$, i.e.\ the lower edge of the noninteracting band in the QPC center lies at $\omega_b = -2\tau + V_g =  -1.5\tau$, c.f.\ Fig.~\ref{fig_model}(a);
screening length \changed{$\chi=5$}. 
%\scrap{, where $a$ denotes the lattice constant of our discretization.} 
This is on the scale of the characteristic length of our QPC, see below; 
magnetic field $B=0$.

The curvature of the central barrier, which sets the characteristic energy scale of the QPC, is then given by
$\Omega_x = 2\sqrt{V_g \tau}/N \approx 0.05\tau$. 
Likewise, the characteristic QPC length scale is given by $l_x=a\sqrt{\tau / \Omega_x} \approx 5a$, \changed{where $a$ denotes the lattice constant of our discretization.}
Moreover, if not otherwise specified, we will use the following set of interaction parameters.
Onsite interaction: $L_U=0$, $U \equiv U_0=0.7\tau=3.2\sqrt{\Omega_x \tau}$. 
\changed{These values were also used in \cite{Schimmel2017}.
%In particular, it was shown that this interaction strength is sufficient to reproduce the 0.7-physics appearing in the magnetic field dependence of the conductance at zero temperature. 
We remark that this onsite interaction strength is close to its maximal value that can be used before the fRG flow breaks down. } 
Finite-ranged interaction: $L_U=3$, $U_0=0.5\tau=2.3\sqrt{\Omega_x \tau}$, $U_1 = 0.3\tau = 1.4\sqrt{\Omega_x \tau}$. 
\changed{These parameters are chosen in such a way that (i) $L_U > l_x/(2a)$ i.e.\ a particle on the top of the QPC barrier can interact with a particle outside of the QPC center, whose width is set by the characteristic length $l_x$. 
(ii) The strength of the onsite term $U_0=0.5$ in \eqref{Yukawa} is chosen to be slightly smaller than that for the pure onsite interaction with $U=0.7$, in order to compensate for the finite extent of the interaction. 
The strength of the offsite interaction is chosen in an ad hoc fashion as $U_1=0.3$ which, as we will see, is large enough to lead to a noticeable impact on the conductance behavior.
In the end of Sec.~\ref{dynamic_feedback_length} we take a very brief look on how the conductance changes with (i) increasing interaction range $L_U$ and (ii) when varying the overall interaction strength while keeping the ration $U_0/U_1$ fixed.
A systematic study of the conductance dependence on the detailed form of the interaction is, however, beyond the scope of this paper.  
  } 
The resulting barrier and interaction forms \changed{for this choice of parameters} are shown in Fig.~\ref{fig_model}. 

\changed{Primarily, we} are interested in the form of the first conductance step that occurs when the QPC opens up, right after pinch-off. 
To vary the effective barrier height, we vary the chemical potential $\mu$ instead of the gate voltage $V_g$, as done in experiments.
This has the advantage that the curvature $\Omega_x$ of the central barrier does not change during the conductance step. 
All observed changes during the step therefore depend only on the energetic distance of the chemical potential to the barrier top, i.e. on the Fermi energy at the central site, $\epsilon_F = \mu - \omega_b$.

\section{Method}
\label{Method}
In order to compute the conductance from the described model, we use a second-order truncated Keldysh fRG (in a similar fashion as described in \cite{Schimmel2017}). 
However, in order to treat finite-ranged interactions we extend the scheme used there, applying an eCLA-approximation, as described in \cite{Weidinger2017}. 

This section is divided into three parts. 
Sec.~\ref{Keldysh_fRG_setup} summarizes the general Keldysh fRG approach to the QPC model \eqref{hamiltonian}. 
Since this general approach is the same as in \cite{Schimmel2017}, we provide only a brief description and just state the most important relations. 
In Sec.~\ref{subsection_eCLA}, we describe the combination of Keldysh- and eCLA fRG in detail, %\scrap{give} 
\changed{discuss} the resulting flow equations and comment on symmetries of the involved quantities.  
Finally, in Sec.~\ref{conductance_computation} we discuss how to obtain the conductance from our fRG data, using the approach presented in \cite{Heyder2017}.

\subsection{Keldysh fRG setup} \label{Keldysh_fRG_setup}
\subsubsection{Propagators} \label{sec_propagators}
We implement our fRG flow as hybridization flow \cite{Jakobs2010, Schimmel2017}, by introducing a flow parameter $\Lambda$ into the \changed{retarded} bare propagator which nominally acts as coupling strength between the system sites (including the leads) and an artificial source of dissipation 
\begin{align}
G^R_{0,\Lambda}(\omega) = \frac{1}{\omega - H_0 + \frac{i}{2} \Lambda},
\label{bare_propagator}
\end{align}
where $H_0$ denotes the single-particle part of the Hamiltonian \eqref{hamiltonian}.
\changed{
Via the relations \eqref{propagators_hermitian_conjugation_a} and \eqref{propagator_fdt}, the $\Lambda$ dependency will also enter the advanced and the Keldysh component of the bare propagator.  	
}
In the limit $\Lambda \rightarrow \infty$ which serves as a starting point of the flow, the artifical dissipation renders the model trivial, whereas for $\Lambda \rightarrow 0$ we recover the full bare propagator. 

As usual, before carrying out any numerical calculations, the non-interacting leads can be integrated out analytically \cite{Karrasch2008,Jakobs2010,Bauer2013} and their effect is absorbed into a self-energy contribution $\Sigma_{\text{lead}}$ for the central region given by sites $[-N,\dots,N]$.
This contribution is located at the two ends of the central region and its retarded component is given by \cite{Schimmel2017}
\begin{align}
\Sigma^{R \sigma \Lambda}_{\text{lead} ij}(\omega) &= \frac{1}{2}(\delta_{i,-N}\delta_{j,-N} + \delta_{i,N}\delta_{j,N}) \nonumber \\
                                                   &\times \Big( \omega_\sigma + i \frac{\Lambda}{2} - i \sqrt{4\tau^2 - \big(\omega_\sigma + i \frac{\Lambda}{2}\big)^2} \Big),
\end{align}
with $\omega_\sigma=\omega \changed{-} \frac{\sigma}{2} B$.
% \scrap{
% Its advanced $\Sigma^A$ and Keldysh components $\Sigma^K$ can be determined via the general relation $\Sigma^A = \Sigma^{R\dagger}$ and (since we consider thermal equilibrium) the fluctuation-dissipation theorem (FDT) $\Sigma^K = (1-2n_F) (\Sigma^R - \Sigma^A)$.
% Here, $n_F(\omega) = (1 + e^{(\omega-\mu)/T})^{-1}$ denotes the Fermi distribution with chemical potential $\mu$ and temperature $T$ (Boltzmann constant $k_B=1$ by convention). 
% }
Using this quantity, the \changed{retarded} bare propagator $G^R_{0,\Lambda ij}(\omega)$ with $i,j$ within the central region can be expressed as 
\begin{align}
G^R_{0,\Lambda ij}(\omega) = \Big[\frac{1}{\omega - H^C_0 - \Sigma^{R \sigma \Lambda}_{\text{lead}} + \frac{i}{2} \Lambda} \Big]_{ij},
\label{bare_propagator_central}
\end{align} 
where $H^C_0$ is the part of the single-particle Hamiltonian that lives entirely within the central region.

Using the $\Lambda$ dependent bare propagator \changed{\eqref{bare_propagator_central}}, the \changed{retarded component of the} single-scale propagator can be obtained by 
\begin{align}
S^R(\omega) &= (G G_0^{-1} \partial_\Lambda G_0 G_0^{-1} G)^R_{\changed{\Lambda}} \changed{(\omega)} \nonumber \\
            &= G^R_\Lambda \changed{(\omega)} \Big(-\frac{i}{2} + \partial_\Lambda \Sigma^{R\Lambda}_{\text{lead}}(\omega) \Big) G^R_\Lambda \changed{(\omega)}.
\label{prop_s_ret}
\end{align}
\changed{In order to simplify notation, we will supress the index $\Lambda$ in the following.}

\changed{
For all propagators and the self-energy, the advanced component is the hermitian conjugate of the retarded component and the Keldysh component is its own negative hermitian conjugate, i.e.\ for all $\xi \in \{G_0,G,S,\Sigma\}$ 
we have
\begin{subequations}
\begin{align}
\xi^A = (\xi^R)^\dagger, \label{propagators_hermitian_conjugation_a}\\
\xi^K = -(\xi^K)^\dagger.\label{propagators_hermitian_conjugation_b}
\end{align}
\label{propagators_hermitian_conjugation}
\end{subequations}
Additionally, due to our equilibrium setup, these quantities also fulfill a fluctuation-dissipation theorem (FDT)   
\begin{align}
\xi^K(\omega) = (1 - 2 f(\omega))\Big(\xi^R(\omega) - \xi^A(\omega)\Big).
\label{propagator_fdt}
\end{align}
Here, $f(\omega) = (1 + e^{(\omega-\mu)/T})^{-1}$ denotes the Fermi distribution with chemical potential $\mu$ and temperature $T$ (Boltzmann constant $k_B=1$ by convention). 

For further use, we also note that using Keldysh indices $\in \{1,2\}$ we have
\begin{align}
G^R = G^{21}, \ G^A = G^{12}, \ G^K = G^{22}.
\label{propagators_keldysh_indices}
\end{align}
Here and in the following sections, we use the common notation, where ``$2$'' indicates the classical and ``$1$'' the quantum component, c.f.\ \cite{Jakobs2009,Jakobs2010a}.
}

\subsubsection{Keldysh and frequency structure of the vertex}
We arange the Keldysh structure \changed{of the two-particle vertex} according to the convention \cite{Jakobs2009,Jakobs2010a} 
\changed{
\begin{align}
\gamma^{\alpha \beta | \gamma \delta} =
\begin{pmatrix} (11|11) & (11|21) & (11|12) & (11|22) \\              
                (21|11) & (21|21) & (21|12) & (21|22) \\                
                (12|11) & (12|21) & (12|12) & (12|22) \\                
                (22|11) & (22|21) & (22|12) & (22|22)    \end{pmatrix},
\end{align}
}
where $\alpha,\beta,\gamma,\delta \in \changed{ \{1,2 \}}$ denote Keldysh indices. 
 
Furthermore, we use a channel decomposition,
\begin{align}
\gamma(\omega_1',\omega_2' | \omega_1, \omega_2) &\approx \bar{\nu} + \varphi^P(\Pi) + \varphi^X(X) 
                                                 + \varphi^D(\Delta), 
\label{channel_decomposition}
\end{align}
with the bosonic frequencies given by 
\changed{
\begin{subequations}
\begin{align}
\Pi &= \omega_1 + \omega_2 = \omega_1' + \omega_2', \\
X &=\omega_2 - \omega_1' = \omega_2' - \omega_1, \\ 
\Delta &= \omega_2 - \omega_2'= \omega_1' - \omega .
\end{align}
\label{bosonic_frequencies}
\end{subequations}
}
\changed{
The quantity $\bar{\nu}$ denotes the bare vertex whose Keldysh structure reads \cite{Jakobs2010} 
\begin{align}
\bar{\nu}^{\alpha_1' \alpha_2' | \alpha_1 \alpha_2} = \tfrac{1}{2} \bar{v}
\begin{pmatrix} 0 & 1 & 1 & 0 \\              
                1 & 0 & 0 & 1 \\                
                1 & 0 & 0 & 1 \\                
                0 & 1 & 0 & 1    \end{pmatrix}.
\label{barevertex_keldysh_structure}
\end{align}
The spin and spatial dependence of the antisymmetrized quantity $\bar{v}$ is given by 
\begin{align}
\bar{v}^{\sigma_1' \sigma_2' | \sigma_1 \sigma_2}_{j_1' j_2' | j_1 j_2} &= 
\delta_{j_1' j_1} \delta_{j_2' j_2}
\delta_{\sigma_1' \sigma_1} \delta_{\sigma_2' \sigma_2} 
U^{\sigma_1 \sigma_2}_{j_1 j_2} \nonumber \\
&- 
\delta_{j_1' j_2} \delta_{j_2' j_1}
\delta_{\sigma_1' \sigma_2} \delta_{\sigma_2' \sigma_1} 
U^{\sigma_1 \sigma_2}_{j_1 j_2}, 
\label{vbar}
\end{align}
with
\begin{align}
U^{\sigma_1 \sigma_2}_{j_1 j_2} = \begin{cases} 0, \text{ if } j_1=j_2 \text{ and } \sigma_1 = \sigma_2 \\ U_{j_1 j_2}, \text{ else.} \end{cases} 
\label{Usigma}
\end{align}
 
The quantitites $\varphi^P(\Pi)$, $\varphi^X(X)$, $\varphi^D(\Delta)$ denote the contributions of the respective channels.
}
Using general symmetries of the vertex, as well as additional (approximate) symmetries introduced by our chosen approximation of the fRG equations, it can be shown that  \cite{Jakobs2010, Schimmel2017} the form of the resulting Keldysh structure depends on the individual channel and is given by 
\begin{subequations}
\begin{align}
\varphi^P(\Pi) = \begin{pmatrix} 0 & d^P & d^P & 0 \\
                              a^P & b^P & b^P & a^P \\ 
                              a^P & b^P & b^P & a^P \\ 
                              0 & d^P & d^P & 0 \end{pmatrix} (\Pi),
\end{align}
\begin{align}
\varphi^X(X) = \begin{pmatrix} 0 & d^X & a^X & \changed{b^X} \\
                              a^X & b^X & 0 & d^X \\ 
                              d^X & 0 & b^X & a^X \\ 
                              b^X & a^X & d^X & 0 \end{pmatrix} (\Chi),
\end{align}
and
\begin{align}
\varphi^D(\Delta) = \begin{pmatrix} 0 & a^D & d^D & b^D \\
                              a^D & 0 & b^D & d^D \\ 
                              d^D & b^D & 0 & a^D \\ 
                              b^D & d^D & a^D & 0 \end{pmatrix} (\Delta).
\end{align}
\label{keldysh_components}
\end{subequations}
Furthermore, including frequency, spin and spatial structure one finds that these components are not all independent but fullfill additional symmetry relations (see App.~\ref{app_sym}).  
In %\scrap{particular} 
thermal equilibrium, it is possible to express all $d$-components via the complex conjugate of $a$-components, see \eqref{ad_connection}. 
Additionally, %\scrap{in thermal equilibrium} 
the components of the vertex fulfill a FDT \cite{Jakobs2009,Jakobs2010a},
\begin{subequations}
\begin{align}
b^P &=  2i \Imp(a^P) \coth \Big( \Big(\frac{\Pi}{2} - \mu \Big)/T \Big), \\
b^X &= -2i \Imp(a^X) \coth \Big( \frac{\Chi}{2T} \Big), \\
b^D &=  2i \Imp(a^D) \coth \Big( \frac{\Delta}{2T} \Big), \label{vertex_d_fdt}
\end{align}
\label{vertex_fdts}
\end{subequations}
leaving the $a$-components as the only independent part of the Keldysh structure. As a final remark, we emphasize that in the chosen convention $a^P(\Pi)$ and $a^D(\Delta)$ are both retarded, whereas $a^X(\changed{X})$ is advanced \cite{Jakobs2009,Jakobs2010a}. 

\subsubsection{Frequency parametrization} \label{subsection_frequency}
We now briefly explain the nature of our chosen frequency parametrization and introduce some notations that will be useful in the subsequent sections. 
Here again, we closely follow the method described in \cite{Schimmel2017}, therefore we refer the interested reader to its extensive supplement material.  
Since we are working in the Keldysh formalism, both the fermionic frequencies in the propagators and self-energy as well as the bosonic frequencies of the vertices are continuous real numbers and one cannot formally distinguish them (as one does in the finite temperature Matsubara formalism). 
%\scrap{Therefore, the general structure of the frequency parametrization for propagators and vertices is the same.} 
\changed{
For our numerical treatment, we use two different frequency parametrizations. 

The first one discretizes the state of the system, i.e.\ self-energy and vertices, with $N_\text{freq}$ underlying frequency points. 
Since both computation time and allocated memory depend crucially on $N_\text{freq}$, this number should be chosen with care. 
For the explicit implementation of the grid, we proceed then as follows.
}
Within the energy window $[-4\tau,4\tau]$, corresponding to twice the band width introduced through our tight-binding leads, we choose a linear discretization, outside of this window we use an exponentially-spaced discretization scheme.
Of the number $N_\text{freq}$ of total frequency points, we use roughly $2/3$ of them within and $1/3$ outside of the linear window. 
In addition to this underlying grid, we add a number of extra frequencies, which depend upon whether we want to use \changed{the grid for the 
%\scrap{fermionic propagators and} 
self-energy,} the P-channel, or the XD-channel contribution of the vertex.   
The idea here is that for each of those cases there is a frequency window of special physical interest.
\changed{For the %\scrap{fermionic propagators/}
self-energy,} this window is around the chemical potential, and for the vertex channels around the so-called feedback frequency, which equals $2\mu$ in the P- and $0$ in the X-channel.  
In each of these cases we add one extra frequency point at each of these special frequencies.
Additionally, in the case of finite temperature, $N_T$ frequencies are added to resolve a frequency window \changed{$[-5T,5T]$} of width \changed{$10T$} around the special frequencies. 
We use $N_\text{freq} \sim \changed{1490}$ and $N_T =10$ and have verified that \changed{
our results are converged w.r.t.\ these two parameters. 
While the number of base grid frequencies $N_\text{freq} \sim 1490$ was already used in \cite{Schimmel2017}, the chosen number of additional frequencies ($\sim 100$) to resolve the temperature window in \cite{Schimmel2017} was much higher than our $N_T=10$. 
Our comparatively low choice of this number is due to the fact, that for our study $N_T$ affects the numerical cost much more than for \cite{Schimmel2017}, due to the inclusion of the long-range part of the vertex around the feedback frequencies, see Sec.~\ref{dynamic_feedback_length}.  
However, even with the choice $N_T=10$, our data is still reasonably converged w.r.t.\ $N_T$, see App.~\ref{app_Convergence_N_T}. 
}
We use the following notation for the frequency parametrization:
We denote the total number of frequency points by $N_{f}$ for the fermionic grid, and by $N_A$ with $A \in \{P,X\}$, for the bosonic P-, and XD-channel grid.
We denote the respective frequency grids by $\omega_{f} = \{ \omega_n \}_{0 \leq n \leq N_{f}}$ and  $\Omega^{A} = \{ \Omega^A_n \}_{0 \leq n \leq N_{A}}$.
We introduce the notation $\Omega^A_f$ for the feedback frequency of the bosonic channels, i.e.\ $\Omega^P_f = 2\mu$ and $\Omega^X_f = 0$.
Moreover, we denote %\scrap{the index of the chemical potential $n_f$ and} 
\changed{the index of the feedback frequency by $n_{A}$.}
Thus, \changed{we have %\scrap{$\omega_{n_f}=\mu$,} 
$\Omega^P_{n_{P}} = 2\mu$} and $\Omega^X_{n_{X}} = 0$.

\changed{
A second frequency parametrization is utilized to discretize the propagators $G$ and $S$.
In a precomputation step, taking place before the evaluation of the r.h.s.\ of the fRG flow equations, we evaluate $G$ and $S$ on a very fine grid of approximately $N_\text{pre}\sim30000$ frequency points, using linear interpolation of the self-energy. 
Whenever a propagator within the r.h.s.\ of the flow has to be evaluated at a given arbitrary frequency (not necessarily a grid frequency) we use its linearly interpolated value obtained from this fine frequency grid.  
Concretely, this evaluation always occurs as part of a frequency integration over an internal fermionic frequency $\omega$, see \eqref{self_flow}, \eqref{bubbles_elementar} below.
Due to the matrix inversion involved in the computation of a propagator from the self-energy, the precomputation method is much faster than computing the propagators separately for each internal frequency occuring in the frequency integration.  
Compared to the time the actual evaluation of the r.h.s.\ takes, the time spent for this precomputation is negligible.
In order to facilitate the integration, we employ a frequency substitution (see discussion in Sec.~\ref{implementational_details}).    
In all our computations, the fine propagator grid was chosen as a uniform grid in this substituted frequency space. 

At the end of this subsection, we summarize the introduced parameters for our frequency grids in Tab.~\ref{tab_freq}. 
The specified values for the number of frequencies will be used for all subsequent calculations, except in App.~\ref{app_Convergence_N_T}, where we discuss the convergence behavior w.r.t.\ $N_T$.} 
\begin{table}
\caption{\changed{Summary of parameters for frequency grids.}}
\begin{center}
	\changed{
    \begin{tabular}{ | c | p{6.7cm} |}
    \hline
    Parameter & Description  \\ \hline
	$N_\text{freq} \sim 1490$
	& Number of basic grid frequencies for self-energy and vertices. 
	%In our case $N_\text{freq} \sim 1490$. 
	\\ 
	\hline
	%$N_T$ 
	$N_T = 10$
	& Additional frequencies in the temperature window $[-5T,5T]$ around the feedback frequencies for the respective vertex channels and the chemical potential for the self-energy. 
	%In our case $N_T = 10$. 
	\\ 
	\hline
	$\Omega^A$ & Resulting frequency grid for channel $A \in \{P,X \}$. 
	\\ 
	\hline
	%$N_A$ 
	$N_A \sim 1500$ & Total number of frequencies in $\Omega^A$. 
	%In our case $N^A \sim 1500$. 
	\\ 
    \hline
	$\Omega^A_f$ & Feedback frequency of channel $A$: \\ 
	& $\Omega^P_f=2\mu$, $\Omega^X_f=0$. 
	\\ 
    \hline
	$n_A$ & Index of the feedback frequency of channel $A$: \\
	& $\Omega^P_{n_P}=2\mu$, $\Omega^X_{n_X}=0$. \\ 
    \hline
	$\omega_f$ & Resulting frequency grid for self-energy. \\ 
    \hline
	%$N_f$ 
	$N_f \sim 1500$
	& Total number of frequencies in $\omega_f$. 
	%In our case $N_f \sim 1500$.
	\\ 
    \hline
    $N_\text{pre} \sim 30000$
	%$N_\text{pre}$ 
	& Total number of frequencies in the fine propagator grid. 
	%In our case $N_\text{pre} \sim 30000$.
	\\ 
    \hline
    \end{tabular}
	}
\end{center}
	\label{tab_freq}
\end{table}

\subsection{Extended Coupled Ladder Approximation} \label{subsection_eCLA}
\subsubsection{Spatial short indices and simple eCLA} \label{simple_eCLA}
Having summarized the general Keldysh setup in the previous subsection, we are now in the position to formulate the fRG flow equations using a variation of the eCLA-Method \cite{Weidinger2017}. 
For this, we first introduce spatial 
``short'' indices $l,k$ \changed{and ``long'' indices $j,i$} , parameterizing the spatial structure of the vertices, as: 
\begin{subequations}
\begin{align}
\changed{(}a^P\changed{)}^{l k}_{ji}(\Pi)    &= a^{P}_{j(j+l) | i(i+k)}(\Pi), \\
\changed{(}a^X\changed{)}^{l k}_{ji}(\Chi)   &= a^{X}_{j(i+k) | i(j+l)}(\Chi),\\
\changed{(}a^D\changed{)}^{l k}_{ji}(\Delta) &= a^{D}_{j(i+k) | (j+l)i}(\Delta). 
\end{align}
\label{definition_short_indices}
\end{subequations}
Since the treatment of the full spatial structure of the vertex is numerically too costly,
the eCLA scheme restricts the range of the short indices $l,k$ by introducing the feedback-length $L$, with $|l|,|k|\leq L$. 
\changed{
The range of the corresponding long indices $j,i$ is dependent on $l,k$,
respectively, since we require that both $j,i$ and $j+l$, $i+k$ lie within the central region, i.e.\   
\begin{subequations}
\begin{align}
\max(-N,-N-l) \leq j \leq \min(N,N-l) \label{ranges_long_indices_j} \\
\max(-N,-N-k) \leq i \leq \min(N,N-k). \label{ranges_long_indices_i}
\end{align}
\label{ranges_long_indices}
\end{subequations}
}
Generically, the feedback length $L$ should be chosen at least as great as the range of the bare interaction $L_U$ ($L \geq L_U$), such that the spatial structure of all vertex components generated in second-order of the bare interaction can be represented. 
In practical applications, we view $L$ as an internal numerical parameter in which convergence should be reached. 
For example, in case of a QPC with onsite-interactions \cite{Weidinger2017} and a static implementation of the eCLA, convergence in the conductance was achieved for $L \approx l_x\changed{/a}$, where $l_x$ is the characteristic length of the QPC. 

However, in this form the eCLA is still too costly to be implemented in a dynamic Keldysh setup, due to the large number of frequencies needed to resolve sharp structures on the real frequency axis:  
A straightforward parameterization with \changed{$N_{P}=N_{X} \sim 1500$} bosonic frequencies, as was chosen in \cite{Schimmel2017}, is numerically not possible if we want to take a feedback length $L$ into account that is at least of the order of the characteristic QPC length $L \approx l_x/a \sim 5$, where $a$ is the lattice spacing of the spatial discretization.
For this reason, we have to further refine our eCLA scheme, see Sec.~\ref{dynamic_feedback_length} below.
However, to do this efficiently, we first take a look at the \changed{structure} of the Keldysh-fRG flow equations. 
%\scrap{formulated in the short-index notation introduced in Eqs.~\eqref{definition_short_indices}}.     

\subsubsection{Flow equations} \label{sec_flow_equations}
\changed{
In this subsection, we state the general form of the flow equations for self-energy and two-particle vertex. 
In order to get a feeling for their general structure, we will not write down their full index dependencies, but rather focus on the important aspects. 
In App.~\ref{app_explicit_flow} the flow equations are then given with their full index structure. 

Due to the equilibrium symmetries of self-energy and vertex (a thorough discussion of these is included in App.~\ref{app_sym}), we only have to compute the flow of $\Sigma^R$ and the $a$ components of the vertex. 
In our presentation here, we will first present the Keldysh and frequency structure and suppress spin and spatial indices. 
For the self-energy this flow takes the following form
\begin{align}
&\partial_\Lambda  \Sigma^R(\omega) \ \widehat{=} \int d\omega' \Big\{ S^R(\omega')\Big[b^X(\omega' - \omega) + b^D(0) \Big] \nonumber \\
                                                        +&S^A(\omega')\Big[b^P(\omega' + \omega) + b^D(0) \Big] \nonumber \\
                                                        +&S^K(\omega')\Big[ \tfrac{1}{2} \bar{v} + a^P(\omega' + \omega) + a^X(\omega' - \omega) + a^D(0) \Big] \Big\}, 
\label{self_flow}
\end{align}
where we have written ``$\,\widehat{=}\,$'' instead of ``$\,=\,$'' in order to indicate that we suppressed a (non trivial) spin and spatial structure. 
Via the relations (\ref{propagators_hermitian_conjugation_a},\ref{propagator_fdt}), $S^A$ and $S^K$ can be expressed through $S^R$ and the $b$ components can be expressed through the $a$ components using the vertex FDTs \eqref{vertex_fdts}. 
Therefore, the flow of the retarded self-energy can be expressed solely through $S^R$ and the $a$ compontents of the vertex. 
By splitting \eqref{self_flow} into a static and a dynamic part, its spatial structure can be expressed using only two pairs of short-long indices $(j,l)$ and $(i,k)$, see App.~\ref{app_explicit_flow}. 
For each combination of those one has to compute an internal frequency integral. Therefore the computational effort for the self-energy scales like $(2N+1)^2 (2L+1)^2$.  

The flow of the $a$ components of the vertex is of the general structure
\begin{align}
\partial_\Lambda a^A(\Omega) \ \widehat{=} \ \tilde{a}^A(\Omega) \, I^A(\Omega) \, \tilde{a}^A(\Omega), 
\label{a_flow}
\end{align}
with $A \in \{P,X,D\}$ and correspondingly $\Omega \in \{\Pi,X,\Delta \}$. 
Again we have suppressed spatial und spin indices, for details see App.~\ref{app_explicit_flow}.
In \eqref{a_flow} , the tilded quantities are given by 
\begin{align}
\tilde{a} \ \widehat{=} \ \tfrac{1}{2} \bar{v} + a^A + \phi^B + \phi^C, 
\end{align}
where $\phi^B,\phi^C$ denotes the static feedback from the other two channels, which is chosen as in \cite{Jakobs2009,Jakobs2010a,Schimmel2017}, namely 
$\phi^P = a^P(2\mu)$,
$\phi^X = a^X(0)$,
$\phi^D = a^D(0)$.
The main effort in the vertex flow goes into the computation of the bubble quantities $I^A(\Omega)$, which contain the internal frequency integration. 
Suppressing spatial and spin structure, these bubbles are of the form 
\begin{subequations}
\begin{align}
I^P  \ &= \ (\tilde{I}^{pp})^{22|21} + (\tilde{I}^{pp})^{22|12} \label{channel_bubbles_P} \\ 
I^X  \ &= \  (\tilde{I}^{ph})^{22|12} + (\tilde{I}^{ph})^{21|22} \label{channel_bubbles_X}\\
I^D  \ &= \  -\Big[(\tilde{I}^{ph})^{22|21} + (\tilde{I}^{ph})^{12|22}\Big], \label{channel_bubbles_D}
\end{align}
\label{channel_bubbles}
\end{subequations}
with
\begin{widetext}
\begin{subequations}
\begin{align}
(\tilde{I}^{pp})^{\alpha_1' \alpha_2' | \alpha_1 \alpha_2}(\Pi) &\widehat{=} \frac{i}{2\pi} \int d\omega \Big[ S^{\alpha_1' \alpha_1}(\omega) G^{\alpha_2' \alpha_2}(\Pi - \omega) + [S \leftrightarrow G] \Big], \label{bubbles_elementar_pp}\\
(\tilde{I}^{ph})^{\alpha_1' \alpha_2' | \alpha_1 \alpha_2}(X) &\widehat{=} \frac{i}{2\pi} \int d\omega \Big[ S^{\alpha_1' \alpha_1}(\omega) G^{\alpha_2' \alpha_2}(\omega + X) + [S \leftrightarrow G] \Big],\label{bubbles_elementar_ph}
\end{align}
\label{bubbles_elementar}
\end{subequations}
\end{widetext}
and the Keldysh convention \eqref{propagators_keldysh_indices}.

Let us now take a look at the spatial structure of \eqref{a_flow}.
We have already seen in \eqref{definition_short_indices} that $(a^A)^{lk}_{ji}$ has a blockmatrix structure in position space, with two pairs of short and long indices $(l,j)$ and $(k,i)$.
The same is true for the bubble quantities $(I^A)^{lk}_{ji}$. 
If we introduce the block-matrix multiplication in spacial indices
\begin{align}
[A \cdot B]^{lk}_{ji}=A^{l k_1}_{j i_1} B^{k_1 k}_{i_1 i}, 
\label{block_matrix_multiplication}
\end{align}
the multiplications appearing between the different factors in \eqref{a_flow} are all of this blockmatrix type, although for the D-channel some factors are to be transposed. 
For details see App.~\ref{app_explicit_flow}.
In our regime of parameters, the bottleneck in computation time is not the blockmatrix multiplications in \eqref{a_flow} but rather the computation of the bubbles \eqref{bubbles_elementar}. 
Therefore, as for the self-energy, the leading contribution to computation time for the r.h.s.\ of the vertex flow scales as $(2N+1)^2 (2L+1)^2$.

After having specified the flow-equations, the last piece missing to determine the flow completely are the initial conditions. 
For a finite but large $\Lambda_\text{ini}$ (in practice $\Lambda_\text{ini}=10^5\tau$) they are given by \cite{Jakobs2009,Jakobs2010a}
\begin{align}
\Sigma^{R \sigma \Lambda_{\text{ini}} }_{ij}(\omega) &= \frac{1}{2} \sum_{k\tau} \changed{\bar{v}}^{\sigma \tau | \sigma \tau}_{i k | j k}, \\
a^{P \Lambda_{\text{ini}}} &= a^{X \Lambda_{\text{ini}}} = a^{D \Lambda_{\text{ini}}} = 0.
\end{align}
}

\subsubsection{Bubble symmetries}
\label{sec_bubble_symmetries}
Since the evaluation of the bubble integrals in \eqref{bubbles_elementar} will be the most expensive part of the fRG flow, 
% \scrap{let us discuss symmetries that can be exploited in the\changed{ir} computation.}   
% \scrap{
% For this we introduce a short-index notation for the spatial indices of the bubble (note that we use the same convention for both the $pp$- and $ph$-contributions):
% }
\changed{we briefly comment on simplifications occurring due to symmetry relations of the bubbles.
While we refer the interested reader again to App.~\ref{app_explicit_flow} for details, it turns out that we only need to compute two Keldysh components of the bubbles \eqref{bubbles_elementar}, namely
\begin{subequations}
\begin{align}
I^{pp} = (\tilde{I}^{pp})^{22|21}, \label{bubble_keldysh_components_pp}\\ 
I^{ph} = (\tilde{I}^{ph})^{22|12}. \label{bubble_keldysh_components_ph}
\end{align}   
\label{bubble_keldysh_components}
\end{subequations}
}
\changed{Thus} generically, we have to compute $8$ integrals of the type given in \eqref{bubbles_elementar}, namely $(I^{pp})^{\sigma \tau}$ and $(I^{ph})^{\sigma \tau}$ for all possible spin combinations of $\sigma,\tau = \pm \uparrow, \downarrow$.
%\scrap{Due to} 
\changed{In} thermal equilibrium, the propagators $G$ and $S$ \changed{for our system} are symmetric in position space (see discussion in App.~\ref{app_General_symmetries}), i.e. 
\begin{subequations}
\begin{align}
G^{\sigma}_{ji}(\omega) &= G^{\sigma}_{ij}(\omega) \label{prop_transp_sym_a} \\
S^{\sigma}_{ji}(\omega) &= S^{\sigma}_{ij}(\omega). \label{prop_transp_sym_b}
\end{align}
\label{propagator_transposition_symmetry}
\end{subequations}
Due to this  property, the bubbles satisfy
\begin{align}
I^{lk}_{ji} = I^{kl}_{ij}.
\end{align}
This implies that we only have to compute the components of the bubble with $k\geq l$, and for $l=k$ only the components with $i\geq j$. 

A further great simplification occurs in the case of zero magnetic field: Here we only need to compute the two integrals $(I^{pp})^{\uparrow \uparrow}$ and $(I^{ph})^{\downarrow \downarrow}$.

\subsubsection{Dynamic feedback length}
\label{dynamic_feedback_length}
\begin{figure}
   \includegraphics[scale=0.5]{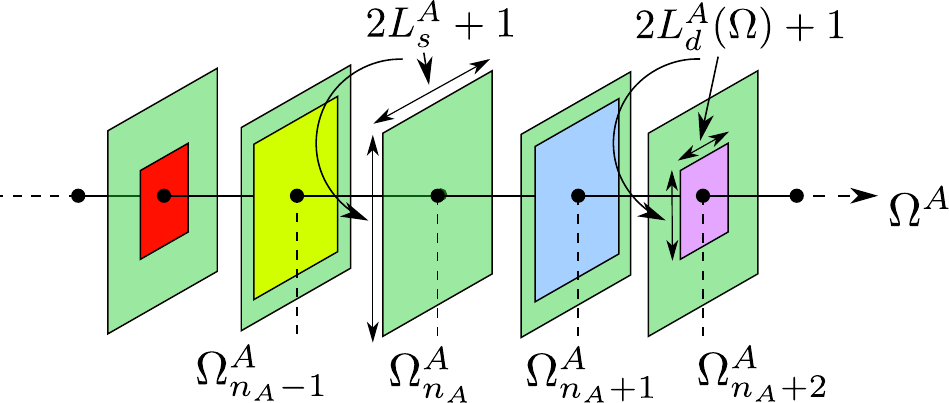} 
   \caption{Illustration of the dynamic feedback length $L^A(\Omega)$.
	        The vertex contribution at the feedback frequency is depicted in green, contributions at other frequencies are shown in different colors.
	        Note that for frequencies $\Omega \neq \Omega^A_{n_A}$, vertex contributions beyond the dynamic feedback length $L^A(\Omega)$ but within the static feedback length $L^A_s$ are replaced by the green feedback contributions.}
\label{freq_str_static_feedback}
\end{figure}

Now that we have obtained the fRG equations, we can proceed to tackle the problem identified in Sec.~\ref{simple_eCLA}:
the huge numerical cost arising from the combination of high frequency resolution in the vertex ($N_A \sim 1500$) with a finite feedback length on the scale of the QPC length $L \sim l_x\changed{/a} \sim 5$ sites.    
Our Ansatz to overcome this challenge is to introduce for each channel $A$ two individual feedback lengths, a static one, $L^A_s$, and a dynamic one, $L^A(\Omega)$, which depends on the bosonic frequency $\Omega$ of the respective channel and decreases with increasing difference between $\Omega$ and the feedback frequency $\Omega^A_f$.
We choose these feedback lengths in such a way that $L^A(\Omega) \leq L^A_s $ for all $\Omega$ and that at the feedback frequency $L^A(\Omega^A_f) = L^A_s$ holds.  
Our strategy is now the following:
For each dynamic block-matrix quantity $M^A \in \{a^A, I^A\}$,
we compute the components $M^{A lk}_{ji}(\Omega)$ (we suppress spin indices in this subsection) only for the spatial and frequency grid points for which $|l|,|k|\leq L^A(\Omega)$ holds.
Thus, using the dynamic feedback length, we can restrict the numerical effort to obtain and store the spatial structure of these quantities for each frequency individually. 
On the other hand, if we have to evaluate $M^A$ in a computation for a short-index  $|l|$ or $|k|$ greater than $L^A(\Omega)$, we apply the following rule:
\begin{align}
M^{A lk}_{ji}(\Omega) = \begin{cases} 0, \text{if } |l|>L^A_s \text{ or } |k| > L^A_s \\ M^{A lk}_{ji}(\Omega^A_f), \text{else.} \end{cases}
\label{static_replacement}
\end{align}  
Thus, if we do not have the dynamic value for a combination of short indices $(l,k)$ available, we replace it, if possible, by the corresponding value at the feedback frequency.
Otherwise we have to set it to zero.  
A schematic illustration of this procedure is given in Fig.~\ref{freq_str_static_feedback}. 
In the special case $L^A(\Omega) = L$ for all $\Omega$ and $A \in \{P,X\}$, we recover the simple eCLA scheme described in \ref{simple_eCLA}.

Using this extended scheme, we are able to include a long-range contribution at physically important frequencies, namely the ones around the feedback frequencies $\Pi=2\mu$ in the P- and $\Chi=0$ in the XD-channel. Those frequencies can be shown to have the biggest contribution to low-energy observables like the linear conductance. 
A short argument for this can be found in App.~\ref{importance_feedback}.   
For all other frequencies we can treat the long-range feedback in a static manner, similar to the treatment in \cite{Weidinger2017}: 
Everytime, we have to evaluate the long-range contribution at one of those frequencies, we will simply replace it by its value at the feedback frequency of the respective channel.   

\changed{This approximation is admittedly quite crude. However, 
note that many previous treatments that were even cruder, e.g.\ treating the vertex only statically altogether, still led to reasonable results. 
In this sense, our semi-static treatment should be understood as the next step on the way to a more quantitatively reliable method.
The approximation could be improved by not using the values at the 
feedback frequency, but the values at the edge of the region that was 
parametrized in detail when going beyond that region. However,
in our view, such a more refined treatment would only be warranted if at
the same time one also refrained from making the channel decomposition
of the vertex. Recall that the channel decomposition tracks only a
single frequency argument per channel and evaluates the contributions
from the other two channels only at the feedback frequency. The errors
incurred in this manner seem to be comparable to the ones incurred by
the approximation of Eq.~\eqref{static_replacement}. A more sophisticated
parametrization of the frequency dependence is left for future work.}

The remaining question is how to choose the frequency dependence of the dynamic feedback length $L^A(\Omega)$. 
Note that generically, for this scheme to be formally exact in second-order in the bare interaction, $L^A(\Omega)$ would have to be chosen greater than $L_U$ for all frequencies in the grid. 
However, this is exactly the situation we want to avoid with this construction:
The hope is that the relevant (low energy) physics can already be captured with a (much) smaller dynamic feedback length when evaluating quantities away from their respective feedback frequencies. 
Thus our goal is to choose a sequence of parameterizations $L^A_n(\Omega)$ that
(a) formally converges pointwise to $2N$ (the maximal value of the feedback length): $\lim_{n \rightarrow \infty} L^A_n(\Omega) = 2N$, and 
(b) achieves a much quicker convergence than the formal one in low-energy observables, yielding an efficient low-energy description. 
In principle, one is free to choose such a sequence in any way one likes. 
In this work, we use a very simple treatment, with a parameterization $L^A(\Omega)$ characterized by only two numerical integer parameters, $L \geq 0$ and $N_L \geq 0$, where $2N_L+1$ sets the window of frequencies around $\Omega^A_f$ within which we treat the long-ranged part of the vertex dynamically.
In fact, we here choose these two parameters channel independent and refer to $L$ as \emph{the} feedback length and $N_L$ as the number of long-range frequencies.  
Physically, the contributions around the feedback frequency $\Omega^{f}_A$ are most important, i.e.\ there it is important to resolve the long-range structure in frequency.
We call this frequency range $\Theta^f_A$ and choose it in a symmetric fashion around the feedback frequency via $\Theta^f_A = [\Omega^A_{n_A - N_L}, \Omega^A_{n_A + N_L}]$. 
Therefore we set $L^A(\Omega)=L$ for all $\Omega \in \Theta^A_f$.
Away from the feedback frequency, we expect a static treatment of the long-range structure to be acceptable, therefore we set the dynamic feedback length $L^A(\Omega) =0$ for all $\Omega \notin \Theta^A_f$. 
In the limit of large $L$ and $N_L$, we recover the full channel decomposed description of the vertex as given in \eqref{channel_decomposition}. 

Note that for a fixed finite $L>0$, and for all observables that depend only on the low energy properties of the system (like e.g.\ the linear conductance) this method interpolates between two extreme cases:
As discussed above, for a large number of long-range frequencies $N_L$, the results of this method converge to the results obtained without static long-range feedback. 
On the other hand, for $N_L=0$ (i.e.\ the only long-range contributions live at the feedback frequencies) this method still already incorporates the spatial structure of the long-range feedback $L$, even though only statically.
Loosely speaking, this $N_L=0$ case results from the simplest possible combination of the previous dynamic work on Keldysh-fRG \cite{Schimmel2017} and the static eCLA implementation in \cite{Weidinger2017}.   
By further increasing $N_L$, we can deepen the combination between those approaches and create more reliable dynamic results.     

\changed{
At the end of this subsection, we summarized the introduced numerical parameters for the dynamic feedback length in Tab.~\ref{tab_eCLA}.}
\begin{table}
\caption{\changed{Summary of parameters for dynamic feedback length}}
\begin{center}
	\changed{
    \begin{tabular}{ | c | p{6.7cm} |}
    \hline
    Parameter & Description  \\ \hline
	%$N$ & central region consists of sites $[-N,N]$. \\ 
	%\hline
	$L^A(\Omega)$ & Dynamic feedback length. Controls the spatial extent of the vertex that is taken into account at frequency $\Omega$.    \\ 
	\hline
	$L^A_s$ & Static feedback length, $L^A_s = L^A(\Omega^A_{n_A})$. For all other frequencies $\Omega$ we have $L^A(\Omega) \leq L^A_s$.   \\ 
	\hline
	$\Theta^f_A$ & Frequency range around the feedback frequency, for which $L^A(\Omega)$ is non-vanishing. Concretely, $L^A(\Omega)=L$ for $\Omega \in \Theta^f_A$ and zero otherwise.  \\ 
	\hline
	$N_L$ & $2N_L+1$ is the number of frequencies in $\Theta^f_A$. \\
	& Concretely, $\Theta^f_A = [\Omega^A_{n_A - N_L},\Omega^A_{n_A + N_L} ]$. \vspace{0.1cm}\\ 
    \hline
    \end{tabular}
	}
\end{center}
	\label{tab_eCLA}
\end{table}

\subsubsection{Further implementational details} \label{implementational_details}
The coupled system of flow equations \changed{(\ref{self_flow_up},\ref{self_flow_down})} and \changed{\eqref{flow_complete}} was solved with a standard fourth-order Runge-Kutta ODE solver. 
The integration over frequencies on the r.h.s.\ of the flow equations was carried out using Gaussian quadrature with Patterson sets \cite{Patterson1968}.
In order to facilitate the computation, we used a substitution of the real frequency axis to the interval $(-7,7)$, which transforms the integrand in such a way that (integrable) poles are avoided and the integrand becomes finite on the whole interval $(-7,7)$. 
This substitution is a slightly modified version of the one used in \cite{Schimmel2017}, see \cite{Schimmel2017a} for details. 
The most time-consuming part of the calculation is the evaluation of the r.h.s.\ of the flow equations, especially the computation of the bubble integrals in the vertex- \eqref{bubbles_elementar_indices} and self-energy flow (\ref{self_flow_up},\ref{self_flow_down}).   
In order to speed up computation time, we used a hybrid MPI + OMP implementation, parallelizing the computation of the self-energy bubble in external frequencies $\omega$ and the vertex bubbles $I^{lk}(\Omega)$ both in external frequency $\Omega$ and additionally in the short-indices $l,k$.     
Furthermore, we also parallelized the block-matrix multiplication appearing on the r.h.s.\ of the flow in the short-indices $l,k$.

\subsection{Conductance Computation} \label{conductance_computation}
The main observable of interest for us is the linear conductance $g$.
In order to compute it, we use a formula first derived by Oguri \cite{Oguri2001}.
We employ its convenient Keldysh formulation developed in \cite{Heyder2017}, whose notational conventions we have also adopted in this work. 
Within this formulation the conductance $g$ can be expressed as  
\begin{align}
g = g_{1} + g_{2}, 
\label{conductance_1}
\end{align}
with the one-particle contribution
\begin{align}
g_{1} = - \frac{e^2}{h} \int^{\infty}_{-\infty } d\epsilon \, f'(\epsilon) \Tr \big\{ \Gamma^l(\epsilon) G^{R}(\epsilon)\Gamma^r(\epsilon) G^{A}(\epsilon)\big\} \label{op_g} 
\end{align}
and the two-particle contribution $g_{2} = g_{2 \Sigma} + g_{2 \Phi}$, with
\begin{subequations}
\label{tp_g}
\begin{align}
g_{2 \Sigma} &= \frac{2e^2}{h} \int d\epsilon \, f'(\epsilon) \Tr \big\{ \Gamma^l(\epsilon) G^{R}(\epsilon) \operatorname{Im}\Sigma^R(\epsilon) G^{A}(\epsilon)\big\} \label{tp_g1}, \\ 
g_{2 \Phi} &=   \frac{e^2}{h} \int d\epsilon \, f'(\epsilon) \Tr \big\{ \Gamma^l(\epsilon) G^{A}(\epsilon){\tilde\Phi}^l(\epsilon) G^{R}(\epsilon)\big\}. \label{tp_g2}
\end{align}
\end{subequations}
Here, $f'$ denotes the derivative of the Fermi distribution 
%\scrap{$f(\epsilon)= 1/(1+e^{\beta(\epsilon - \mu)})$} 
w.r.t.\ energy $\epsilon$, 
$\Gamma^{r}(\epsilon)_{ij}=\delta_{i N} \delta_{j N} \Gamma(\epsilon)$, $\Gamma^{l}(\epsilon)_{ij}=\delta_{-N i} \delta_{-N j} \Gamma(\epsilon)$,
with $\Gamma(\epsilon) = \changed{\theta(2\tau - |\epsilon|)}\sqrt{4\tau^2 - \epsilon^2}$, are the hybridization functions for the right/left lead, $2\operatorname{Im}\Sigma^R = -i(\Sigma^R - \Sigma^A)$ and
$\tilde \Phi^r(\epsilon)$ is the vertex correction term. This term encodes the direct contribution of the two-particle vertex to the conductance. It is given by (c.f. \cite{Heyder2017}, Eq.~(20))
\begin{align}
(\tilde{\Phi}^{l/r})^{\changed{\sigma_2}}_{\changed{j_2' j_2}}(\epsilon) &= \frac{1}{2\pi i} \int d\changed{\epsilon}' \, \sum_{j_1',j_1} \Big[ G^A(\epsilon') \Gamma^{l/r}(\epsilon') G^R(\epsilon') \Big]^{\changed{\sigma_1}}_{\changed{j_1 j_1'}} \nonumber \\
                               &\phantom{=} \times K_{\changed{j_1' j_2' |j_1 j_2}}^{\changed{\sigma_1 \sigma_2 | \sigma_1 \sigma_2}}(\epsilon,\epsilon',0). \label{definition_phi}
\end{align}
The vertex response part $K^{\changed{\sigma_1 \sigma_2 | \sigma_1 \sigma_2}}_{j_1' j_2' |j_1 j_2}(\epsilon,\epsilon',0)$ can be brought into the form (using the vertex FDTs \eqref{vertex_fdts})
\begin{widetext}
\begin{subequations}
\begin{align}
K^{\sigma \sigma | \sigma \sigma}_{j_1' j_2' | j_1 j_2}(\epsilon,\epsilon',0) &= 2i \Big[ \Imp (a^p)^{ \sigma \sigma(j_2' - j_1')(j_2 - j_1)}_{j_1' j_1}(\epsilon' + \epsilon) f^p(\epsilon,\epsilon')   
                                                                               - \Imp (a^d)^{ \sigma \sigma (j_2 - j_1')(j_2' - j_1)}_{j_1' j_1}(\epsilon' - \epsilon) f^x(\epsilon,\epsilon') \Big] \\ 
K^{\bar\sigma \sigma | \bar\sigma \sigma}_{j_1' j_2' | j_1 j_2}(\epsilon,\epsilon',0) &= 2i \Big[ \Imp (a^p)^{(j_1' - j_2')(j_1 - j_2) \sigma \bar\sigma}_{j_2' j_2}(\epsilon' + \epsilon) f^p(\epsilon,\epsilon')  
                                                                                       - \Imp (a^x)^{(j_1' - j_2)(j_1 - j_2') \sigma \bar\sigma}_{j_2 j_2'}(\epsilon' - \epsilon) f^x(\epsilon,\epsilon') \Big]  \\ 
K^{\sigma \bar\sigma | \sigma \bar\sigma}_{j_1' j_2' | j_1 j_2}(\epsilon,\epsilon',0) &= 2i \Big[ \Imp (a^p)^{(j_2' - j_1')(j_2 - j_1) \sigma \bar\sigma}_{j_1' j_1}(\epsilon' + \epsilon) f^p(\epsilon,\epsilon')  
                                                                                       + \Imp (a^x)^{(j_2 - j_1')(j_2' - j_1) \sigma \bar\sigma}_{j_1' j_1}(\epsilon - \epsilon') f^x(\epsilon,\epsilon') \Big],   
\end{align}
\label{def_K}
\end{subequations}
\end{widetext}
with the functions
$f^p(\epsilon,\epsilon') = 2f(\changed{\epsilon'}) + 2b(\epsilon' + \epsilon - \mu)$ and $f^x(\epsilon,\epsilon')= 2f(\changed{\epsilon'}) + 2b(\epsilon' - \epsilon + \mu)$. Here $b(\epsilon) = 1/(e^{\beta(\epsilon -\mu)}-1)$ denotes the Bose distribution.

\begin{figure}
   \includegraphics[scale=1.0]{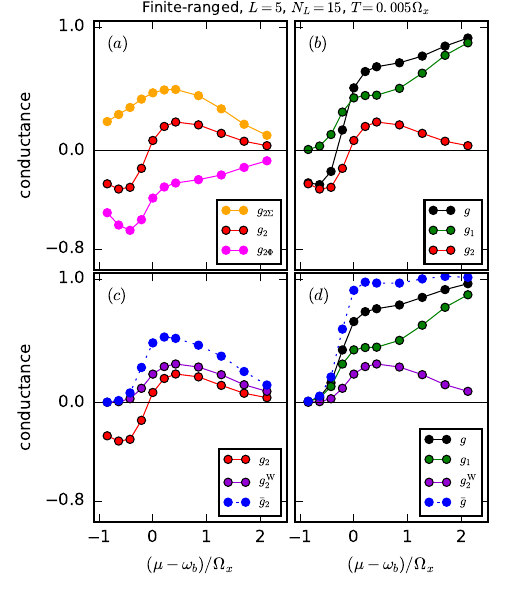} 
	\caption{\small Conductance obtained via straightforward application of formulas (\ref{op_g}-\ref{tp_g}).
(a) Two-particle contributions $g_2 = g_{2\Sigma} + g_{2\Phi}$ [Eq.~\eqref{tp_g}]. 
(b) Single- and two-particle contributions to the total conductance $g=g_1 + g_2$ [Eq.~(\ref{conductance_1}-\ref{tp_g})]. Note that both $g_2$ and $g$ are negative at pinch-off. 
(c) Comparison of $g_2$ %\scrap{and} 
to $g^{\textrm W}_2$ and $\bar{g}_2$; the latter \changed{two go} to zero at pinch-off.
(d) Single-particle and Ward-corrected two-particle contributions to the total conductance $g=g_1 + g^{\textrm W}_2$.
\changed{For comparison we also show $\bar{g} = g_1 + \bar{g}_2$.}} 
\label{fig_two_pos}
\end{figure}

Fig.~\ref{fig_two_pos} shows the resulting conductance for a generic set of parameters. 
Fig.~\ref{fig_two_pos}(a) depicts the two-particle contributions $g_{2}$, $g_{2 \Sigma}$, and $g_{2 \Phi}$. 
In particular, note that for small values of the chemical potential $\mu$, the \changed{total} two-particle contribution becomes negative. 
This carries over to the total conductance, see Fig.~\ref{fig_two_pos}(b): 
At pinch-off, the one particle-contribution $g_1$ vanishes and thus the negative two-particle part $g_2$ leads to a negative conductance $g$. 
This behavior is clearly unphysical, as the total conductance should vanish below pinch-off.
The cause of this problem has to stem from the two major approximations that we applied: The channel decomposition \eqref{channel_decomposition} and the general second-order fRG truncation. 
Especially the latter is known to lead to a violation of the law of current conservation and %\scrap{and} 
Ward identities (see App.~\ref{vio_ward_identities} for a more detailed discussion). 
In particular, the Ward identity 
\begin{align}
\tilde\Phi^{l}(\epsilon) + \tilde\Phi^r(\epsilon) = - 2 \operatorname{Im} \Sigma^R(\epsilon),
\label{ward_identity}
\end{align}
derived in \cite{Heyder2017}, is violated in our approximation scheme,
leading to unphysical results for transport quantities \cite{Schimmel2017a}.
To ameliorate this problem, we replace the vertex contributions $\tilde \Phi^{l/r}$ by ``Ward-corrected'' versions,
\begin{align}
\tilde \Phi^{l/r,\textrm{W}}_{ij} (\epsilon) = \tilde \Phi^{l/r}_{\changed{ij}} (\epsilon)F_{ij}(\epsilon),
\quad
F_{ij}(\epsilon) = \frac{-2 \operatorname{Im} \Sigma^R_{ij}(\epsilon)}{(\tilde\Phi^r+\tilde\Phi^l)_{ij}(\epsilon)}.
\label{ward_symmetrized}
\end{align}
The multiplicative factor $F_{ij}$ nominally equals $1$ if $\tilde \Phi^{l,r}$ satisfy the Ward identity \eqref{ward_identity} with $\operatorname{Im} \Sigma^R$.
If they do not, it by construction ensures that $\tilde \Phi^{l/r,\textrm{W}}$ \textit{do},
\begin{align}
\tilde\Phi^{l,\textrm{W}}(\epsilon) + \tilde\Phi^{r,\textrm{W}}(\epsilon) = - 2 \operatorname{Im} \Sigma^R(\epsilon),
\end{align}
thereby compensating the adverse consequences of the second-order truncation scheme.
(To avoid numerical errors arising from division by very small numbers, we set $F_{ij}(\epsilon) = 1$ whenever its denominator becomes smaller than $10^{-8}$;
the results are not sensitive to the value of this bound.)
The sum of \eqref{tp_g1} and \eqref{tp_g2}, with $\tilde \Phi^l$ replaced by $\tilde \Phi^{l,\textrm{W}}$ in the latter, yields 
\begin{align}
g^{\textrm{W}}_{2} = - \frac{e^2}{h} \int^{\infty}_{-\infty } d\epsilon \, f'(\epsilon) \Tr \big\{ \Gamma^l(\epsilon) G^{A}(\epsilon){\tilde\Phi}^{r,\textrm{W}}(\epsilon) G^{R}(\epsilon)\big\}. \label{tp_g3}
\end{align}
Note that the integrand is proportional to $\tilde\Phi^r$.
This property ensures that the conductance vanishes at pinch-off, as can be seen by the following argument. 
Assume that the QPC is closed, i.e.\ the chemical potential $\mu$ is below the QPC barrier. 
Then in the integral \eqref{tp_g3} only frequencies $\epsilon$ below the QPC barrier contribute, implying that the propagators $G^{R/A}_{ij}(\epsilon)$ are only non-vanishing for spatial indices $i,j$ on the same side of the barrier. 
Therefore, since the hybridization function $\Gamma^l(\epsilon)$ lives on the left side of the system,  
only contributions of $\tilde\Phi^r_{ij}(\epsilon)$ contribute where $i,j$ are on the left side of the barrier.
However, applying the same logic in the definition of $\tilde\Phi^r(\epsilon)$ \eqref{definition_phi}, we see that $\tilde\Phi^r_{ij}(\epsilon)$ is only non-vanishing for $i,j$ on the right side of the barrier.
Therefore, the two-particle part of the conductance vanishes at pinch-off.  
Indeed, this is confirmed by the violet curves in Fig.\ref{fig_two_pos}(c,d), computed using Eq.~\eqref{tp_g3} for $g^{\textrm{W}}_{2}$. 

All conductance results shown in the subsequent sections are obtained using the Ward-corrected two-particle contribution \eqref{tp_g3}. 

Note that if one evokes the Ward identity \eqref{ward_identity} without replacing $\tilde\Phi^{l/r}$ by $\tilde\Phi^{l/r,\textrm{W}}$, the sum of \eqref{tp_g1} and \eqref{tp_g2} yields an expression for $g_{2}$ similar to \eqref{tp_g3}, but containing $\tilde\Phi^r$ instead of $\tilde\Phi^{r,\textrm{W}}$. 
This expression \changed{$\bar{g}_2$}, which corresponds to the second term in Eq.~(23) of \cite{Heyder2017}, also vanishes at pinch-off, \changed{see Fig.~\ref{fig_two_pos}(c,d)}.
However, we believe it to be unreliable when used in conjunction with second-order-truncated fRG, since the latter, as mentioned above, yields results for $\tilde \Phi^{l,r}$ which (in contrast to $\tilde \Phi^{l/r,\textrm{W}}$) violate the Ward identity used for its derivation.
 
\section{Results}
\label{Results}
In this section, we investigate the features one obtains for a QPC with a finite-ranged interaction of the type described in Sec.~\ref{Model}. 
The section is divided into two parts.
In the first part, we present results obtained with a dynamic treatment of the short-range part and a static treatment of the long-range part of the vertex.  
In the second part, both short-range and long-range contributions of the vertex are treated dynamically.

\subsection{Static long-range part}
\begin{figure}
   \includegraphics[scale=1.0]{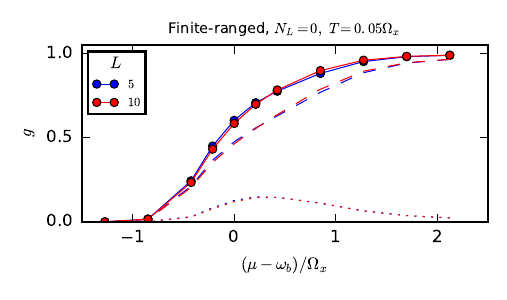} 
   \caption{\small Conductance for large feedback lengths $L=5,10$ (solid curves), computed using a static treatment of the long-ranged part of the vertex, i.e.\ using $N_L=0$. Dashed and dotted curves indicate the one- and two-particle contribution, respectively.  As in the static Matsubara case, we see that $L=5$ is sufficient to achieve convergence.}
\label{fig_conv_in_L}
\end{figure}

The results of this first subsection are obtained \changed{using $N_L=0$, i.e.}\ by a direct combination of the dynamic treatment of the short-range part \cite{Schimmel2017} and the static treatment of long-range part of the vertex \cite{Weidinger2017}.
As discussed in Sec.~\ref{subsection_eCLA}, introducing a finite-ranged interaction necessitates the introduction of the feedback length $L$, measuring the range over which the vertex develops structure during the RG flow. 
In \cite{Weidinger2017}, we have shown that in the static Matsubara setup convergence in $L$ was reached for $L\sim l_x/a$ and $L>L_U$, where $l_x$ is the characteristic QPC length and $L_U$ the range of the interaction.    
In our new Keldysh formulation, this statement remains true. 
As an example, Fig.~\ref{fig_conv_in_L} shows a typical conductance curve for our generic finite-ranged interaction from Sec.~\ref{Model}, computed at a finite temperature $T=0.05\Omega_x$.  
We see that convergence is reached around $L=5 \approx l_x/a$.
In the rest of this work, we always use $L=5$ if not explicitly stated otherwise.

\begin{figure}
   \includegraphics[scale=1.0]{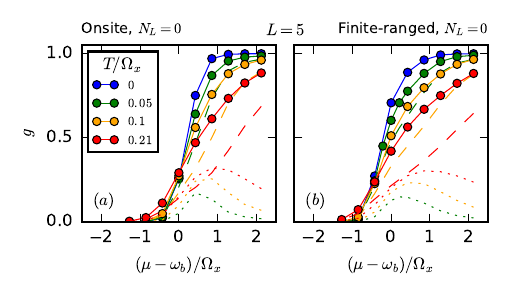} 
   \caption{\small Temperature dependence of the conductance (solid curves) for a model with (a) onsite interations and (b) finite-ranged interactions ($L_U=3$), computed using a feedback length $L=5$ and static long-range part $N_L=0$. Dashed and dotted curves indicate the one- and two-particle contributions, respectively. In the finite-ranged case (b) the conductance shows a slightly stronger flattening in the $0.7$ region than in the onsite case (a). However, the form of the curves is still quite similar.}
\label{fig_stat_temp_dep}
\end{figure}

Having assured the convergence w.r.t.\ the feedback length, we can now compare the implication of finite-ranged interactions on the conductance within a \emph{static} long-range feedback description. 
For this, we compare a typical onsite-interaction model (here we use the same parameters as used in \cite{Schimmel2017}, in particular onsite $U=0.7\tau$) with a model with finite-ranged interactions. 
The form of the interaction is here chosen as introduced in Sec.~\ref{Model}, i.e.\ with a onsite interaction strength $U=0.5\tau$ and exponentially screened offsite components, reaching an interaction range of $L_U=3$. Therefore, a particle in the center of the QPC can directly interact with a particle outside the center, being half the characteristic QPC length away.   
The resulting conductances are shown in Fig.~\ref{fig_stat_temp_dep}.
Fig.~\ref{fig_stat_temp_dep}(a) displays the conductance of the onsite model, which is qualitatively very similar to the one obtained in \cite{Schimmel2017}, even though we here use a finite feedback length $L$. 
It is important to mention that in \cite{Schimmel2017} this onsite interaction strength was chosen as large as possible without causing a failure of convergence for the RG flow. 
However, even in this maximal interaction strength case, no development of a pronounced 0.7-shoulder with increasing temperature was observed.
In Fig.~\ref{fig_stat_temp_dep}(b) we use a finite-ranged interaction.
The only difference compared to part (a) is that the conductance curves are slightly more asymmetric, indicating that due to its finite range, the amount of interaction that can be taken into account with fRG is larger.
However, there is still no pronounced shoulder in the conductance.
In the next subsection, we will see that this changes when taking a dynamic contribution of the long-range part into account. 

\subsection{Dynamic long range part}
\label{subsection_long_range}
\begin{figure}
   \includegraphics[scale=1.0]{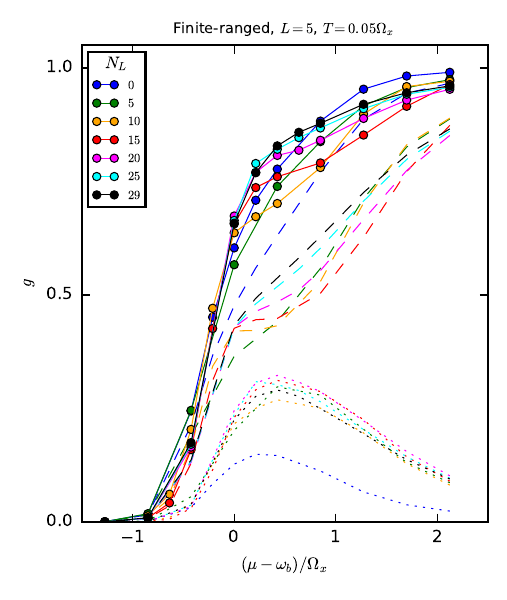} 
   \caption{\small Dependence of the conductance on increasing $N_L$, which controls the width of the frequency window within which the long-ranged part of the vertex is treated dynamically, at finite $L=5$. While, within our numerical resources, convergence in $N_L$ could not be fully reached, finite values of $N_L$ seem to lead to a more prominent 0.7-feature than in the onsite case: This is most pronounced for medium $N_L=10,15$ and still noticeable at large $N_L=25,29$.}
\label{fig_dyn_temp_dep_NL}
\end{figure}

In this section, we will extend our study by treating the long-range part of the vertex dynamically within a certain window of frequencies.
As explained in Sec.~\ref{subsection_eCLA}, this window is controlled numerically by the number, $N_L$, of frequency points around the feedback frequencies \changed{for which the long-range part is} %\scrap{that are} 
taken into account.  
However, there is a caveat: Our frequency parametrization is not strictly uniformly spaced, especially around the feedback frequencies we have to distinguish two scales, c.f.\ Sec.~\ref{subsection_frequency}.
The smaller scale is set by temperature, and we use %\scrap{a number}
\changed{$N_T=10$} %\scrap{of} 
frequencies distributed on that scale around the feedback frequency to resolve the temperature dependence.  
The other relevant scale is set by the curvature $\Omega_x$, 
which is resolved by our underlying equally spaced general frequency grid, introduced in Sec.~\ref{subsection_frequency}. 
Therefore, when we increase $N_L$ up to $\sim 5$ we take only the vertex contribution in a frequency range set by temperature into account. 
A further increase of $N_L$ then begins also to resolve the $\Omega_x$ scale, which sets the scale of the characteristic width of the conductance step.
Concretely, the half-width of the frequency range of the long-range vertex is given by $\Delta \omega = 0.8\Omega_x$ for $N_L=10$ and increases roughly by $0.8\Omega_x$ per additional increase of $5$ in $N_L$. 
Thus, the biggest value $N_L=29$ corresponds to a maximal frequency range of $\Delta\omega=3.8\Omega_x$.
Furthermore, one can show that the leading frequency contribution to the conductance at the chemical potential $\mu$ lies around the feedback frequencies in a range determined by $\epsilon_F = \mu - V_b$ (c.f.\ App.~\ref{importance_feedback}), i.e.\ it is on a scale set by $\Omega_x$.  
Between $N_L=10$ and $N_L=15$, $\Delta \omega$ becomes bigger than $\Omega_x$.
Thus, starting from $N_L=15$, we take all leading frequency contributions into account for values of the chemical potential reaching the shoulder region, c.f.\ Fig.~\ref{fig_dyn_temp_dep_NL}.
 
The dependence of the resulting conductance on $N_L$ for a typical set of parameters is shown in Fig.~\ref{fig_dyn_temp_dep_NL}.  
Although, we were not able to reach completely converged results at our maximal value $N_L=29$ (after which we hit the memory bound of our computational resources),
 there seems to be a persistent feature for large $N_L$:  
Going from $N_L=0$ (the static long-range result from last section) up to finite $N_L=29$, we observe a qualitative difference in the conductance. 
In the second half of the conductance step a shoulder-like structure emerges, resembling the 0.7-anomaly observed at finite temperature in various experiments \cite{Thomas1996,Appleyard2000,Cronenwett2002,Micolich2011,Iqbal2013,Bauer2013}.   
This feature is most pronounced for $N_L=10-15$, when just the leading frequency contribution is taken into account and relaxes somewhat for larger $N_L$. 
However, as we will show below, even for $N_L=29$ the 0.7-feature is still much more prominent than in the onsite case. 

When decomposing the conductance in one- and two-particle contributions (dashed and dotted lines in Fig.~\ref{fig_dyn_temp_dep_NL}), we see that this 0.7-feature comes from two effects:
(i) In the shoulder region, the one-particle part itself exhibits a kink at a conductance value around $g\sim 0.4$. 
This feature is very strongly pronounced for $N_L=10$ and seems to weaken somewhat for larger $N_L$.
Note here that near pinch-off the differences between curves with different $N_L$ are small and become larger starting when $\mu$ reaches the shoulder region. 
This behavior is consistent with our discussion in App.~\ref{importance_feedback}.  
(ii) The two-particle contribution increases steeply from pinch-off towards its maximum in the shoulder region and decreases after that. 
This feature seems to be almost equally pronounced for all large $N_L=15 - \changed{29}$.     
Both of these effects lead to the development of a shoulder-like structure in the conductance. 

Concluding this discussion, we point out another interesting effect. Even if the one- and two-particle parts themselves are still subject to changes in $N_L$, these changes seem to mostly cancel out each other. The resulting conductance seems to be much lesser dependent on $N_L$: Comparing the magenta ($N_L=20$), cyan ($N_L=25$), and black lines ($N_L=29$) in Fig.~\ref{fig_dyn_temp_dep_NL}, the $N_L=29$ data seem almost converged in the shoulder region.
In fact, apart from the precise position of the shoulder, the qualitative shape of all three curves is already very similar. 
Intuitively this effect makes sense: If a particle traverses the QPC and contributes directly to the conductance via the one-particle contribution it is less likely to have given energy to create particle-hole excitations 
which might contribute to the two-particle part of the conductance and vice versa.

In the following, we study the dependence of the 0.7-feature on temperature, interaction range and interaction strength.
For this, we will always compare the onsite interaction result with the finite-ranged results for both the leading frequency case at $N_L=15$, where the 0.7-structure is most pronounced, as well as for the full $N_L=29$ result. 
\begin{center}
\begin{figure*}
   \includegraphics[scale=1.0]{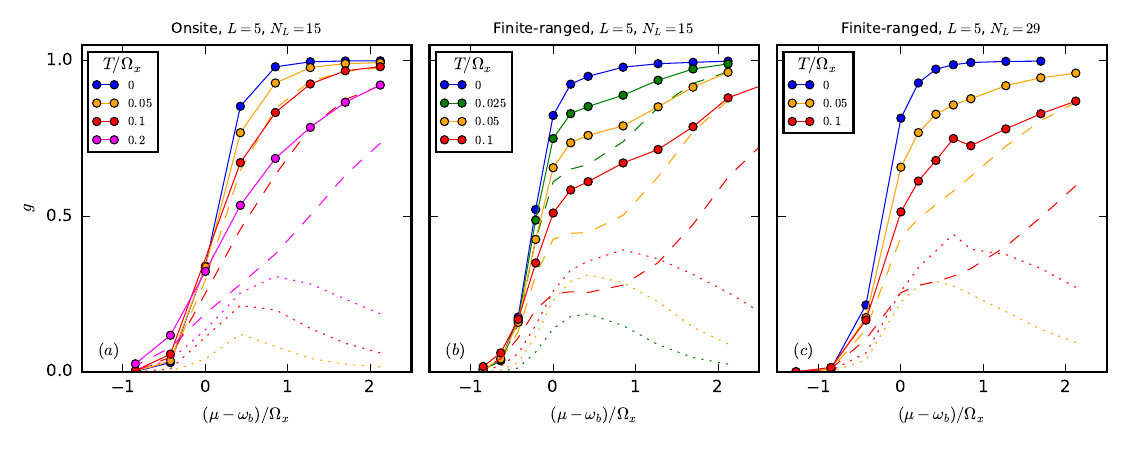} 
   \caption{\small Temperature dependence of the conductance for (a) onsite and (b) finite-ranged interactions with $N_L=15$ and (c) $N_L=29$. In contrast to the onsite case, the finite-ranged conductance shows a much more pronounced 0.7-feature: While for $N_L=15$ in (b) an actual shoulder emerges, the full $N_L=29$ result in \changed{(c)} is still much more asymmetric than the onsite-case.}
\label{fig_dyn_temp_dep}
\end{figure*}
\end{center}

Above we have established the development of a 0.7-shoulder in the finite-ranged interaction model when treating the long-range contributions of the vertex dynamically.
In Fig.~\ref{fig_dyn_temp_dep}, we study how finite-ranged interactions affect the temperature dependence of the conductance. 
We see that the form of the onsite-conductance in Fig.~\ref{fig_dyn_temp_dep}(a) is still the same as in Fig.~\ref{fig_stat_temp_dep}(a,b). 
However, in Fig.~\ref{fig_dyn_temp_dep}(b,c), we see that for finite-ranged interactions increasing temperatures lead to a more and more pronounced 0.7-plateau.
As above, we see that in the $N_L=15$ case the 0.7-feature is most pronounced, however also for $N_L=29$ it is much stronger than in the onsite case.  
In addition to having a different shape, the conductance also depends much more strongly on temperature itself.
We see that finite-ranged interactions, if treated dynamically, have the potential to introduce major changes compared to onsite interactions and are likely to be essential ingredients in the development of a pronounced 0.7-plateau. This finding constitutes the main result of this paper.

While we believe that the qualitative behavior of the conductance is captured correctly within our approach, we still want to comment on two inaccuracies: 
In the $N_L=29$ case, the $T=0.1\Omega_x$ curve exhibits a slight kink in the 0.7-structure, which can be traced back to a peak in the two-particle contribution. 
This is probably an artifact of our method, indicating that for this parameter regime an improvement of the vertex description is in order:   
While it could be that simply a larger value of $N_L$ is needed to converge to a smooth result, it might also be possible that for a more accurate description one would have to improve the vertex treatment alltogether.    
We comment on one possible way to do this below.  
Another problem that we can observe in Fig.~\ref{fig_dyn_temp_dep}(b,c) is a (slight) pinch-off shift to lower chemical potentials, i.e.\ the QPC with finite-ranged interactions opens up earlier than the one with onsite interactions or even the one without interactions. 
This unphysical behavior, an artifact of our method, was also encountered in our earlier work in the Matsubara context \cite{Weidinger2017}.
It will be interesting to see, whether further improvements of the vertex treatment succeed in eliminating this unphysical shift.     

\begin{figure}
   \includegraphics[scale=1.0]{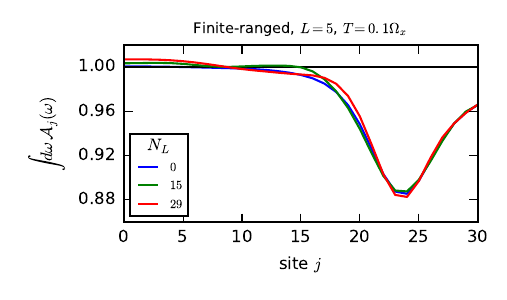} 
   \caption{\small LDOS normalization in the plateau region $(\mu - \omega_b)/T = 0.4$ for finite interaction range for different parameters $N_L$. In the QPC center the normalization condition $\int d\omega \mathcal{A}_j(\omega) = 1$, is satisfied much better than in the flanks.}
\label{fig_ldos_integrated}
\end{figure}

Further insight can be gained by looking at the resulting local density of states (LDOS) of the interacting system.
First of all, this yields an intrinsic consistency check, by inspecting how well the LDOS satisfies the normalization condition $\int d\omega \mathcal{A}_i(\omega) = 1$, see Fig.~\ref{fig_ldos_integrated}.
Note that the normalization condition is relatively well satisfied in the center of the QPC (where the relevant physics for transport happens) and is off in the flanks of the QPC. 
This is somewhat to be expected, since we utilized our numerical resources in such a manner as to best resolve the position and frequency dependence in the center region, i.e.\ for frequencies close to barrier top and chemical potential.
For up to site $15\approx 3l_x/a$ the LDOS normalization is fulfilled well, which is exactly the region of the renormalized flat barrier top, as we will see below. 
Beyond that most of the LDOS contribution sits deeper in the flanks of the QPC away from the barrier top and the region of good resolution. 
Within the region of the barrier top itself, the leading frequency contribution $N_L=15$ seems to be yielding the best results. 

Having checked the LDOS normalization, we next discuss the frequency resolved LDOS structure. 
Fig.~\ref{fig_ldos_comparison} shows the LDOS $\mathcal{A}_i(\omega)$ as a colorplot depending on frequency and site index of the effective QPC barrier. 
Comparing the onsite result (a) to the finite-ranged results (b,c) shows that the latter exhibit a stronger flattening. 
This behavior is qualitatively consistent with our static Matsubara treatment, which also suggested a flatter barrier top for finite-ranged interactions.
Just as the conductance earlier, this indicates again that here more interaction processes are taken into account.
Comparing the two finite-ranged results, the $N_L=15$ result exhibits a stronger van Hove ridge peak than the $N_L=29$ result. 
Applying the rationale developed in \cite{Bauer2013}, this is consistent with the more pronounced 0.7-structure in the conductance in Fig.~\ref{fig_dyn_temp_dep}.

Aside from the form of the renormalized barrier in the 0.7-regime of the conductance step, one can also look at the development of this barrier when varying the chemical potential.
For this we plot in Fig.~\ref{fig_pinning} the LDOS on the middle site $\mathcal{A}_0(\omega)$ as function of frequency and chemical potential, analogously to Fig.~(5) of \cite{Schimmel2017}.   
We see that when the chemical potential (black line) crosses the barrier top $\omega_b$, the van Hove ridge of the interacting LDOS increases with it.
This pinning is much more pronounced for the finite-ranged case [Fig.~\ref{fig_pinning}(b,c)] than for the onsite-case [Fig.~\ref{fig_pinning}(a)].
Again, this indicates the presence of more interaction processes in the case of finite-ranged interactions.

\begin{center}
\begin{figure*}
   \includegraphics[scale=1.0]{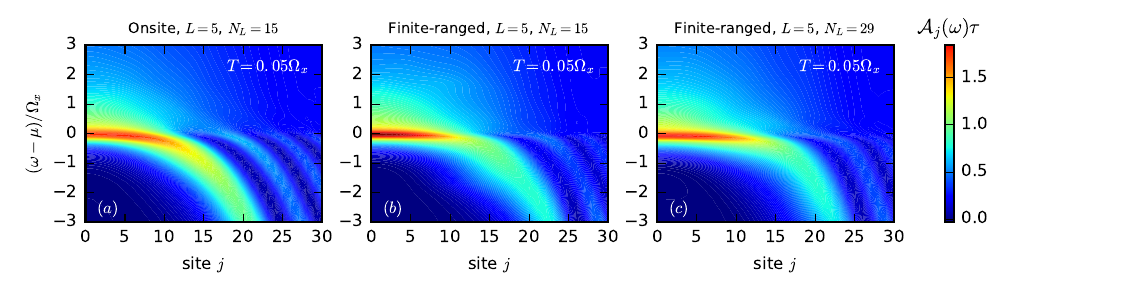} 
   \caption{\small QPC LDOS as function of site and frequency for (a) onsite-, and finite-ranged interactions with (b) $N_L=15$ and (c) $N_L=29$. Note that in (b) and (c) the renormalized barrier top is much flatter than in the onsite case. For the $N_L=15$ case in (b), the LDOS peak in the middle of the QPC is slightly more pronounced than in the $N_L=29$ case (c).}
\label{fig_ldos_comparison}
\end{figure*}
\begin{figure*}
   \includegraphics[scale=1.0]{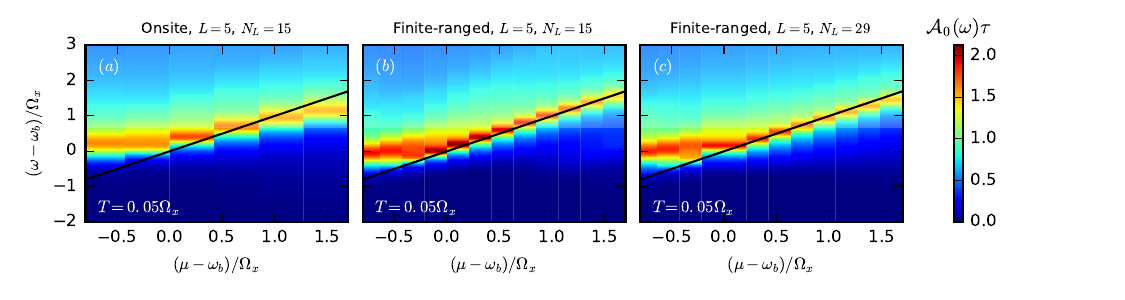} 
   \caption{\small Central QPC LDOS $\mathcal{A}_0(\omega)$ as function of chemical potential and frequency for (a) onsite-, and finite-ranged interactions with (b) $N_L=15$ and (c) $N_L=29$. For finite-ranged interactions the pinning of the van Hove ridge to the chemical potential is much stronger than in the onsite case. Note that in the leading contribution case $N_L=15$, the LDOS is more pronounced than in the full $N_L=29$ result.}
\label{fig_pinning}
\end{figure*}
\end{center}

\begin{figure}
   \includegraphics[scale=1.0]{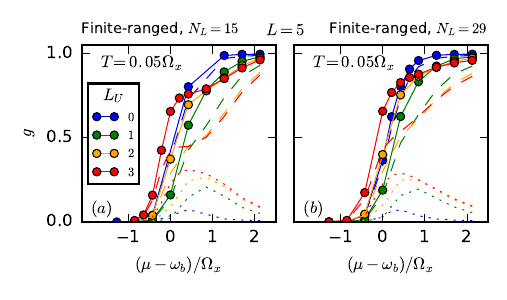} 
   \caption{\small Dependence of the conductance on the interaction range for (a) $N_L=15$ and (b) $N_L=29$. With increasing interaction range the 0.7-feature develops in the conductance step. Note that with increasing $L_U$ the pinch-off of the conductance is shifted to smaller chemical potentials}
\label{fig_inc_range}
\end{figure}

Up to now, we always used the same finite-ranged interaction with an interaction range on the scale of the characteristic length of the QPC and a strength that had been chosen ad hoc.  
A systematic study of how these properties affect the QPC conductance is beyond the scope of this work.
However, in the very last part of this subsection, we will take a first brief look what happens when these parameters are changed. 
Fig.~\ref{fig_inc_range} shows the influence of a variation in the interaction range. 
With increasing interaction cutoff $L_U$, the conductance changes from the onsite $L_U=0$ to the $L_U=3$ results discussed earlier. We see that the 0.7-feature becomes more pronounced, while at the same time the unphysical pinch-off shift mentioned above occurs. 

\begin{figure}
   \includegraphics[scale=1.0]{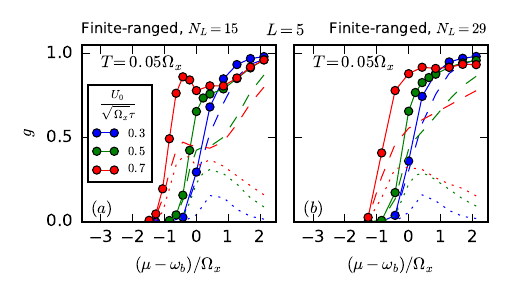} 
   \caption{\small Dependence of the conductance on the interaction strength (a) for the leading frequency contribution $N_L=15$, as well as (b) the full $N_L=29$ contribution. For large interaction strength the 0.7-structure develops an oscillatory feature, more pronounced in (a) but also visible in (b). Note again the unphysical shift to smaller chemical potentials occuring for larger interaction strength.}
\label{fig_inc_strength}
\end{figure}

Fig.~\ref{fig_inc_strength}, instead shows the dependence of the conductance on increasing interaction strength with fixed range $L_U=3$.
Here, we keep the ratio of onsite- and offsite-interaction strength $U_0/U_1=5/3=$ fixed and increase $U_0$ from $0.3\tau$ beyond our usual value $0.5\tau$ to the large value $0.7\tau$.
With increasing interaction strength, the form of the conductance becomes more asymmetric and the 0.7-structure eventually develops a oscillatory feature.  
Similar to the observations discussed above, this is very pronounced for the leading frequency contribution ($N_L=15$) and less visible for $N_L=29$. 
Again the unphysical pinch-off shift in the chemical potential is clearly visible.

\subsection{Further challenges}
\label{further_challenges}
In the data of the previous subsection, we have noticed that for finite-ranged interactions an unphysical shift in the conductance occurs:
The pinch-off is shifted to lower chemical potentials, seeming to imply that the effective QPC barrier gets somehow reduced by finite-ranged interactions. 
%\scrap{This effect is an artefact of the method used.} 
This %\scrap{artefact} 
\changed{effect} was also found to a varying extent in previous fRG work on QPCs \cite{Weidinger2017,Bauer2014,Bauer2013,Heyder2017,Schimmel2017} and is an artefact of our method, presumably our truncation scheme.
%\scrap{is consistent in the sense that we also already observed it in our static Matsubara implementation of the eCLA \cite{Weidinger2017}.} 
\changed{Together with} the other inconsistencies, namely the violation of the Ward identity \eqref{ward_identity} and 
the associated issue that the two-particle contribution to the conductance is negative unless the Ward-correction \eqref{ward_symmetrized} is used, 
this implies that in order to obtain quantitatively reliable results for the conductance one will have to go beyond the channel decomposition \eqref{channel_decomposition}, 
and in general also beyond second-order truncated fRG.    
In particular, a more refined description and treatment of the vertex is required, using not only one but all three %\scrap{bosonic} 
\changed{independent} frequencies.
A possible approach for meeting the latter challenge within the Matsubara formalism is detailed in \cite{Wentzell2016}.
A general improvement of our method could be to combine this efficient vertex treatment with the recently developed multiloop fRG (mfRG) method \cite{Kugler2018a,Kugler2018b,Kugler2018} which provides a natural strategy for going beyond second-order truncated fRG.   
Work in this direction is currently in progress.

\section{Conclusions}
\label{Conclusion}

\jvd{The work reported here had two goals. The first 
was methodological -- advancing fRG methodology by combining long-range
feedback (eCLA) with the Keldysh formalism. The second
goal was phenomenological -- investigating the effect of finite-ranged
interactions on the temperature dependence of the 0.7-anomaly in  QPCs.

Regarding our second goal, the conclusions are encouraging: we find
clear indications that finite-ranged interactions strengthen the 0.7-shoulder in the conductance step at finite temperature. However,
we were unable to fully achieve our first goal: 
 the approximations used (1-loop truncation, channel decomposition of the vertex) are too crude to obtain a fully converged and truly satisfactory fRG treatment of long-range interactions in the Keldysh formalism.  Moreover, we encountered problems arising from the violation of Ward identities. 

 Thus, we conclude that finite-ranged interactions merit further study in the context of the 0.7-anomaly, but more sophisticated methodology is needed to describe them satisfactorily. A promising candidate for further
 studies in this direction would be multi-loop Keldysh-fRG \cite{Kugler2018a,Kugler2018b,Kugler2018}. Work in this direction is currently in progress.  }

% We have applied a Keldysh version of the extended Coupled Ladder Approximation (eCLA) in second-order truncated fRG to a model of a QPC with onsite and finite-ranged interactions.
% Despite problems arising from the violation of Ward identities, 
% we found distinct evidence that finite-ranged interactions are an essential ingredient for the development of a pronounced 0.7-shoulder at finite temperature.

% In order to validate this further, and resolve problems with Ward identities, one could try to use multi-loop fRG \cite{Kugler2018a,Kugler2018b,Kugler2018} in a Keldysh context. Work in this direction is currently in progress.  

\begin{acknowledgments}
We thank Dennis Schimmel, Jan Winkelmann and Edoardo di Napoli for helpful discussions.

Furthermore, we gratefully acknowledge support from the Deutsche Forschungsgemeinschaft through the Cluster of Excellence \emph{Nanosystems Initiative Munich} and Germany's Excellence Strategy-EXC-2111-390814868.

Finally, we gratefully acknowledge the Gauss Centre for Supercomputing e.V. (www.gauss-centre.eu) for supporting this project by providing computing time through the John von Neumann Institute for Computing (NIC) on the GCS Supercomputer JUWELS at J\"ulich Supercomputing Centre (JSC). 
\end{acknowledgments}

\vspace{-0.3cm}
\appendix 

\section*{Appendix}
\changed{
In this appendix, we discuss some more technical aspects of our model and method. 
We begin with summarizing the general symmetries of our system in App.~\ref{app_General_symmetries}. 
These symmetries are exact and do not depend on the channel decomposition or our fRG approximations.   
In App.~\ref{app_sym}, we discuss the implications of these general symmetries on the components \eqref{keldysh_components} of the channel decomposition and count the number of independent components. 
In particular, we use in App.~\ref{app_sym} a more general form of the multiparticle FDTS (\ref{vertex_fdts},\ref{ad_connection}) in the channel decomposition than in previous works \cite{Jakobs2009,Jakobs2010a,Schimmel2017}.
For the interested reader, we have included a derivation of this more general form in App.~\ref{app_derivation_fdts}. 
In App.~\ref{app_explicit_flow}, we show the explicit form of the flow equations from Sec.~\ref{sec_flow_equations}, including the full index structure.
In App.~\ref{importance_feedback}, we discuss the importance of the feedback frequencies for the conductance, and give a justification for our frequency approximation within Sec.~\ref{dynamic_feedback_length} while developing the dynamic feedback length.
In App.~\ref{vio_ward_identities}, we explicitly show the violation of the Ward identity \eqref{ward_identity} for increasing interaction strength. 
Finally, in App.~\ref{app_Convergence_N_T}, we discuss the convergence of our results w.r.t.\ the number of frequencies $N_T$ for which we take a long-range structure of the vertex into account, c.f.\ Sec.~\ref{dynamic_feedback_length}. 
}

\section{General symmetries} \label{app_General_symmetries}
\changed{
In this section, we list the general symmetries that our system introduced in Sec.~\ref{Model} obeys.
The derivation of these symmetry relations can be found in great detail in \cite{Jakobs2009}.
Note that all the symmetries discussed in this section are exact. 
In particular they do \emph{not} depend on the channel decomposition \eqref{channel_decomposition}, or any fRG approximations.
\begin{enumerate}
\item Particle permutation. 
For any permutation $P$ of $(1,\dots,n)$ with sign $(-1)^P$ holds (c.f.\, Eq.~(3.18) in \cite{Jakobs2009}) 
\begin{align}
\xi_{P m' | m } = \xi_{m' |P m } = (-1)^P \xi_{m' | m},
\end{align}
where $\xi \in \{G,\gamma\}$ is either a multi-particle Green's or vertex function and $m=(m_1,\dots,m_n)$ is a multi-particle index, with  $m_k=(\omega_k,\alpha_k,q_k)$ consisting of frequency $\omega_k$, Keldysh index $\alpha_k$, and site and spin index $q_k=(i_k,\sigma_k)$. 
\item Complex conjugation.
For $\xi \in \{G,\gamma\}$ holds (c.f.\, Eq.~(3.24) in \cite{Jakobs2009})
\begin{align}
\xi^{\alpha' | \alpha}_{q' | q}(\omega'| \omega)^* = (-1)^{z_\xi + \sum_k (\alpha_k' + \alpha_k)} \xi^{\alpha | \alpha'}_{q | q'}(\omega | \omega'), \label{gen_cc} 
\end{align}
with $z_G = n$ and $z_\gamma=1$.
For further reference, we also state the equivalent relation of \eqref{gen_cc} on the double time contour (i.e.\ before rotating to Keldysh space)
\begin{align}
\xi^{j'| j}_{q'| q}(t'|t)^* = (-1)^{z_\xi} \xi^{\bar{j} | \bar{j}'}_{q | q'}(t,t'),
\label{gen_cc_dt}
\end{align}
where $j,j' \in \{+,-\}^n$ are multi-particle indices on the double time contour ($+$: forward branch, $-$: backward branch), and $\bar{j}= -j$.

\item Thermal equilibrium and time reversal.
In thermal equilibrium, our system obeys the general Kubo-Martin-Schwinger (KMS) condition \cite{Kubo1957,Martin1959,Kubo1966}, which leads to the relation (c.f.\ Eq.~(3.52) in \cite{Jakobs2009})
\begin{align}
e^{\beta \Delta^{j|j'}(\omega | \omega')} G^{j|j'}_{q | q'}(\omega | \omega') = (-1)^{m^{j | j'}} \tilde{G}^{\bar{j} \bar{j}'}_{q | q'}(\omega | \omega'), \label{kms_con}
\end{align}
with
\begin{align}
m^{j'| j} = \sum_{k: j_k = +} 1 - \sum_{k: j_k'=+} 1,
\end{align}
and 
\begin{align}
\Delta^{j'| j}(\omega' | \omega) = \sum_{k: j_k=+} (\omega_k - \mu) - \sum_{k:j_k'=+}(\omega_k' - \mu).
\end{align}
The tilded Green's function $\tilde{G}$ in \eqref{kms_con} is defined as the normal Green's function $G$, however with anti-time ordering on the forward- and time orderinng on the backward branch, see Eq.~(3.16) in \cite{Jakobs2009}.  
In the single-particle case, $\tilde{G}$ can be expressed simply in terms of $G$ via the relation (c.f. Eq.~(3.17) in \cite{Jakobs2009})
\begin{align}
\tilde{G}^{j'| j}_{q'| q}(\omega' | \omega) = G^{\bar{j} \bar{j}'}_{q'| q}(\omega' | \omega). \label{tilde_G_sp} 
\end{align}
Combining \eqref{kms_con} with \eqref{tilde_G_sp} and rotating to Keldysh space (we follow the convention in \cite{Jakobs2009}, see \eqref{keldysh_rotation}) yields the single-particle FDTs \eqref{propagator_fdt}. 

Additionally to the KMS conditions, thermal equilibrium also implies the following time reversal behavior for multi-particle Green's functions (c.f.\ Eq.~(3.71) in \cite{Jakobs2009}) 
\begin{align}
\tilde{G}^{j | j'}_{q | q'}(\omega | \omega') = G^{\bar{j}' | \bar{j}}_{\tilde{q}' | \tilde{q}}(\omega' |\omega) \Big|_{\tilde{H}}. \label{gen_time_reversal}
\end{align}
Here, $\tilde{q} = \Theta q$ denote the time reversed basis states, where $\Theta$ is the anti-unitary time reversal operator 
\begin{align}
\Theta |i,\sigma \rangle = e^{i \frac{\pi}{2} \sum_k \sigma_k} |i,\bar{\sigma} \rangle, 
\end{align}  
with $\bar{\sigma}$ denoting the opposite spin of $\sigma \in \{+,-\}^n$.
Note that the propagator on the r.h.s.\ of \eqref{gen_time_reversal} has to be evaluated using the time reversed Hamiltonian $\tilde{H}=\Theta H \Theta^\dagger$. 

The Eqs.~\eqref{kms_con} (relating $G$ and $\tilde{G}$) and \eqref{time_reversal} (relating $\tilde{G}$ and $G|_{\tilde{H}}$) are general equilibrium properties. 
Our specific system exhibits additionally a special form of time-reversal symmetry, that will allow us to relate $G$ and $\tilde{G}$:
%Additionally to fulfilling the KMS conditions, our system exhibits in equilibrium also a special form of time-reversal symmetry. 
For the components of the propagators evaluated in the basis $\{| q \rangle \}$ with $|q\rangle=|i,\sigma \rangle$ holds (see Eq.~(3.80) in \cite{Jakobs2009})
\begin{align}
G^{j j'}_{q q'}(t,t') = G^{j j'}_{\tilde{q} \tilde{q}'}(t,t') \Big|_{\tilde{H}}.
\label{time_reversal}
\end{align}
Although our system is more general than the ones considered in \cite{Jakobs2009}, the proof that \eqref{time_reversal} holds for our specific choice of the basis $\{|q\rangle\}$ can be done completely analogously to the one in \cite{Jakobs2009}, pp.~60-61. 
For details, see \cite{Weidinger2020}. 
We remark that for \eqref{time_reversal} to hold, the Hamiltonian \eqref{hamiltonian} does \emph{not} have to be time reversal invariant itself, in particular \eqref{time_reversal} also holds for finite magentic field.

Using \eqref{time_reversal}, we can obtain two more important symmetry relations. 
In the single-particle case, combining \eqref{time_reversal} with \eqref{gen_time_reversal} and \eqref{tilde_G_sp} yields  
\begin{align}
G^{j'|j}_{q' q}(\omega' | \omega) = G^{j' | j}_{q | q'}(\omega | \omega'). 
\label{g_transposition_sym}
\end{align}
Since, in our system, $G$ is diagonal in spin and frequency, this implies that the spatial transposition symmetry \eqref{prop_transp_sym_a} and by extension also \eqref{prop_transp_sym_b}.

In the multiparticle case, one can combine \eqref{time_reversal} with \eqref{gen_time_reversal} and \eqref{kms_con} to obtain after transformation to Keldysh space a FDT for $G$. 
An analog relation holds for the vertex $\gamma$, making it possible to express this multi-particle FDTs for $\xi \in \{G,\gamma\}$ in the compact form (see Eqs.~(3.104,3.106) in \cite{Jakobs2009})
\begin{widetext}
\begin{subequations}
\begin{align}
\Rep \xi^{j'|j}_{\epsilon_{\xi}^{j'| j}}(\omega' | \omega) &= - \Big[1 - 2 f\Big(\Delta^{j'|j}(\omega'|\omega) + \mu\Big) \Big] \Rep \xi^{j' | j}_{-\epsilon_{\xi}^{j'|j}}(\omega'| \omega), \\
\Imp \xi^{j'|j}_{-\epsilon_{\xi}^{j'| j}}(\omega' | \omega)&= - \Big[1 - 2 f\Big(\Delta^{j'|j}(\omega'|\omega) + \mu\Big) \Big] \Imp \xi^{j' | j}_{\epsilon_{\xi}^{j'|j}}(\omega'| \omega),
\end{align}
\label{gamma_fdt}
\end{subequations}
\end{widetext}
where 
\begin{align}
\epsilon_\xi^{j'| j} = (-1)^{1 + n_\xi + m^{j'| j}},  
\end{align}
and for given $\epsilon = \pm 1$
\begin{align}
\gamma^{j'| j}_\epsilon = \sum_{\stackrel{\alpha',\alpha}{(-1)^{\sum_k (\alpha_k' + \alpha_k)}=\epsilon}} D^{j' | \alpha'} \gamma^{\alpha'| \alpha} (D^{-1})^{\alpha | j}, 
\label{gamma_epsilon}
\end{align}
with the Keldysh rotation
\begin{subequations}
\begin{align}
D^{-|1} &= D^{\pm | 2} = \frac{1}{\sqrt{2}}, \\ 
D^{+|1} &= - \frac{1}{\sqrt{2}}.
\end{align}
\label{keldysh_rotation}
\end{subequations}
%Furthermore,
%\begin{align}
%\epsilon_\gamma^{j'| j} = (-1)^{m^{j'| j}}, \\ 
%\epsilon_G^{j'| j} = (-1)^{1+n+m^{j'| j}}, \\ 
%m^{j'| j} = \sum_{k: j_k = +} 1 - \sum_{k: j_k'=+} 1,
%\end{align}
%and 
%\begin{align}
%\Delta^{j'| j}(\omega' | \omega) = \sum_{k: j_k=+} (\omega_k - \mu) - \sum_{k:j_k'=+}(\omega_k' - \mu).
%\end{align}

%For our present work, \eqref{time_reversal} can be used to derive three additional restrictions on our propagators and vertices.  
%Applied to the single-particle case, \eqref{time_reversal} leads to (i) the spatial transposition symmetry of the propagators \eqref{propagator_transposition_symmetry}. 
%In the multi-particle case, \eqref{time_reversal} can be combined with the KMS condition of thermal equilibrium to yield a multi-particle FDT relation, see Eq.~(3.106) of \cite{Jakobs2009}.   
\end{enumerate}
}

\section{Symmetries of vertex components}
\label{app_sym}
In this section, we discuss the symmetries of the vertex components \changed{$\varphi^P$, $\varphi^X$, $\varphi^D$} of Eq.~\eqref{keldysh_components}. 
%This symmetries arise from the complex conjugation and particle exchange symmetry inherent in the definition of the vertex, as well as thermal equilibrium and \changed{a special form of time-reversal symmetry that our system obeys in thermal equilibrium \cite{Jakobs2009}. 
This symmetries arise \changed{ from the general vertex symmetries discussed in App.~\ref{app_General_symmetries}. 
%The latter is a time reversal symmetry relation that holds for our specific representation (i.e.\ it relies on our choice of the single-particle basis $q=|i,\sigma \rangle$) of the propagators of our system . 
%We discuss these special equilibrium properties in more detail in subsection \ref{equilibrium_case}, below. 
We first take a look at the general (i.e.\ not necessarily equilibrium) symmetries in App.~\ref{app_general_case},
and discuss special equilibrium properties in more detail in App.~\ref{equilibrium_case}, 
where we also comment on additional symmetries arising in the case of zero magnetic field or a parity-symmetric model.    
}

%In contour space (i.e.\ before performing the Keldysh rotation) it takes the form (see \cite{Jakobs2009}, p. (?) ) 
%\begin{align}
%G^{\mathbf{r r'}}_{\mathbf{q q'}}(\mathbf{t,t'}) = G^{\mathbf{r r'}}_{\mathbf{\tilde{q} \tilde{q}'}}(\mathbf{t,t'}) \Big|_{\tilde{H}},
%\label{time_reversal}
%\end{align}
%where the bold indices are multiindices, i.e.\ $\mathbf{t} = (t_1, \dots, t_n)$, etc.
%Here, $\mathbf{r,r'}$ denote indices on the double time contour and $\tilde{\mathbf{q}} = \Theta q$ the time reversed basis states  
%Although our system is more general than the ones considered in \cite{Jakobs2009}, the proof that \eqref{} holds for our specific choice of the single particle basis can be done completely analogously to the one in \cite{Jakobs2009}, pp. (?). 
%For details, see \cite{}. 
%We remark that for \eqref{time_reversal} to hold, the Hamiltonian \eqref{} does \emph{not} have to be time reversal invariant itself, in particular \eqref{time_reversal} also holds for finite magentic field.

\subsection{General \changed{case}} \label{app_general_case}
\begin{figure*}
   \includegraphics[scale=0.75]{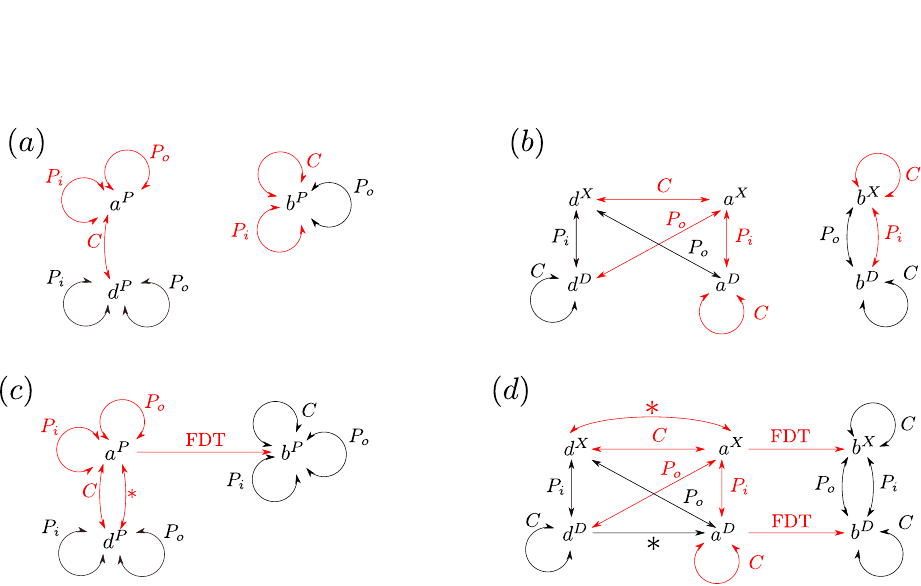} 
   \caption{\small Graphical representation of the symmetry relations for the P-channel (a,c) and XD-channel (b,d). The first row (a,b) depicts the general symmetries for the non-equilibrium case, the second row (c,d) depicts the symmetries for the special case of thermal equilibrium. For each subfigure, the red colored symmetries are an example for an independent subset.}
\label{fig_all_symmetries}
\end{figure*}

Using general vertex properties and the channel decomposition of 2nd-order truncated fRG, one obtains various relations for the vertex components in \eqref{keldysh_components} (c.f. e.g. \cite{Jakobs2009,Jakobs2010a}). 
Fig.~\ref{fig_all_symmetries}(a,b) depicts how those symmetries relate the different components. 
We use the notation:
\begin{itemize}
\item $P_i$: Exchange of incoming particles: \\ 
             $\changed{\varphi}_{\beta_1' \beta_2' | \beta_1 \beta_2} \stackrel{P_i}{\rightarrow} \changed{-\varphi}_{\beta_1' \beta_2' | \beta_2 \beta_1}$,
\item $P_o$: Exchange of outgoing particles: \\
             $\changed{\varphi}_{\beta_1' \beta_2' | \beta_1 \beta_2} \stackrel{P_o}{\rightarrow} \changed{-\varphi}_{\beta_2' \beta_1' | \beta_1 \beta_2}$,
\item $C$: Vertex conjugation: \\
             $\changed{\varphi}_{\beta_1' \beta_2' | \beta_1 \beta_2} \stackrel{C}{\rightarrow} \changed{(-1)^{1 + \sum_k \alpha_k' + \alpha_k}\varphi}^*_{\beta_1 \beta_2 | \beta_1' \beta_2'}$.
\end{itemize} 
Here $\beta=(\alpha,\omega,j,\sigma)$ are \changed{composite}-indices, comprised of Keldysh index, frequency, spatial site and spin.
Each of these three symmetries is depicted by an arrow, connecting related vertex components.
Therefore each of the components is connected via three solid arrows to other components or itself.
The symmetries obey the general relations
\begin{align}
P_i^2 = P_o^2 = C^2 = 1, \nonumber \\
[P_o,P_i] = 0, \nonumber \\
C P_i = P_o C. \label{symmetry_relations}
\end{align}
This implies that not all the relations between the various vertex components are independent, i.e.\ that they can not be expressed via each other. However, one can always find an independent subset of relations. In Fig.~\ref{fig_all_symmetries}, an example for such an independent subset is given by the relations colored red. 
Expressed as equations, this independent subset takes the form
\begin{align}
(a^P)^{\sigma_1' \sigma_2' | \sigma_1 \sigma_2}_{j_1' j_2' | j_1 j_2}(\Pi) &\stackrel{P_o}{=} -(a^P)^{\sigma_2' \sigma_1' | \sigma_1 \sigma_2}_{j_2' j_1' | j_1 j_2}(\Pi), \\
                                                                           &\stackrel{P_i}{=} -(a^P)^{\sigma_1' \sigma_2' | \sigma_2 \sigma_1}_{j_1' j_2' | j_2 j_1}(\Pi), \\
                                                                           &\stackrel{C}{=}    (d^{P*})^{\sigma_1 \sigma_2 | \sigma_1' \sigma_2'}_{j_1 j_2 | j_1' j_2'}(\Pi).
\end{align} 

\begin{align}
(b^P)^{\sigma_1' \sigma_2' | \sigma_1 \sigma_2}_{j_1' j_2' | j_1 j_2}(\Pi) &\stackrel{P_o}{=} -(b^P)^{\sigma_2' \sigma_1' | \sigma_1 \sigma_2}_{j_2' j_1' | j_1 j_2}(\Pi), \\
                                                                           &\stackrel{C}{=}   -(b^{P*})^{\sigma_1 \sigma_2 | \sigma_1' \sigma_2'}_{j_1 j_2 | j_1' j_2'}(\Pi).
\end{align} 

\begin{align}
(a^X)^{\sigma_1' \sigma_2' | \sigma_1 \sigma_2}_{j_1' j_2' | j_1 j_2}(\Chi) &\stackrel{\changed{P_o}}{=} -(\changed{d}^D)^{\sigma_2' \sigma_1' | \sigma_1 \sigma_2}_{j_2' j_1' | j_1 j_2}(\Chi), \\
                                                                            &\stackrel{\changed{P_i}}{=} -(\changed{a}^D)^{\sigma_1' \sigma_2' | \sigma_2 \sigma_1}_{j_1' j_2' | j_2 j_1}(\changed{-}\Chi), \\
                                                                            &\stackrel{C}{=}    (d^{X*})^{\sigma_1 \sigma_2 | \sigma_1' \sigma_2'}_{j_1 j_2 | j_1' j_2'}(\Chi).
\end{align}
\begin{align}
(b^X)^{\sigma_1' \sigma_2' | \sigma_1 \sigma_2}_{j_1' j_2' | j_1 j_2}(\Chi) &\stackrel{P_i}{=} -(b^D)^{\sigma_1' \sigma_2' | \sigma_2 \sigma_1}_{j_1' j_2' | j_2 j_1}(-\Chi), \\
                                                                            &\stackrel{C}{=}   \changed{-}(b^{X*})^{\sigma_1 \sigma_2 | \sigma_1' \sigma_2'}_{j_1 j_2 | j_1' j_2'}(\Chi).
\end{align}

\begin{align}
(a^D)^{\sigma_1' \sigma_2' | \sigma_1 \sigma_2}_{j_1' j_2' | j_1 j_2}(\Delta) \stackrel{C}{=} (a^{D*})^{\sigma_1 \sigma_2 | \sigma_1' \sigma_2'}_{j_1 j_2 | j_1' j_2'}(-\Delta).
\label{ad_freq_sym}
\end{align}

\subsection{Equilibrium case}
\label{equilibrium_case}
%If the system under consideration is in thermal equilibrium, as is the case in this work, we gain additional restrictions \cite{Jakobs2009,Jakobs2010a}, namely the \changed{propagator} FDTs \eqref{}. \eqref{vertex_fdts}, relating $b$'s and $a$'s, as well as the relation
%If the system under consideration is in thermal equilibrium, as is the case in this work, \changed{the propagator FDTs \eqref{propagator_fdt} hold.
\changed{
Besides the generic single-particle FDTs \eqref{propagator_fdt}, which are a generic property of any equilibrium system, the multiparticle relation \eqref{gamma_fdt} holds due to the special form of time-reversal symmetry \eqref{time_reversal} that our system obeys. 
Applying this multi-particle relation to our channel decomposition, we obtain two properties for our vertex quantities, namely 
(ii) the vertex FDTs from \eqref{vertex_fdts}, as well as (iii) the relation
}
% \eqref{vertex_fdts}, relating $b$'s and $a$'s, as well as the relation
\begin{align}
a^* = d,
\label{ad_connection}
\end{align}
 which holds for all channels. 
\changed{
Since especially the relations \eqref{vertex_d_fdt} and \eqref{ad_connection} have (to our knowledge) not been stated in this generality before, we give a short derivation for the interested reader in App~\ref{app_derivation_fdts}.    
}

In Fig.~\ref{fig_all_symmetries}, the symmetries containing 
%\scrap{these}
 \changed{the additional equilibrium symmetry} relations are depicted in panels (c,d). 
In the following, we will restrict our discussion to this equilibrium case.
Then, for finite magnetic field, we have $7$ independent components in spin space:
\begin{subequations}
\begin{align}
\label{spin_components_begin}
(a^P)^{\sigma \sigma} &:= (a^P)^{\sigma \sigma | \sigma \sigma}, \ \sigma=\uparrow, \downarrow, \\
(a^P)^{\uparrow \downarrow} &:= (a^P)^{\uparrow \downarrow | \uparrow \downarrow}, \\
(a^X)^{\uparrow \downarrow} &:= (a^X)^{\uparrow \downarrow | \uparrow \downarrow}, \\
(a^D)^{\sigma \sigma} &:= (a^D)^{\sigma \sigma | \sigma \sigma}, \ \sigma=\uparrow, \downarrow, \\
(a^D)^{\uparrow \downarrow} &:= (a^D)^{\uparrow \downarrow | \uparrow \downarrow}. 
\end{align}
\label{spin_components}
\end{subequations}

The remaining task is to determine the symmetries of these quantities in position and frequency space and to identify the independent components.   
This process can be illustrated again via the symmetry diagrams shown in Fig.~\ref{fig_all_symmetries}.
We are now looking for a complete subset of independent symmetry operations that do not change the channel or spin configuration, i.e. that do not mix the quantities introduced in \eqref{spin_components}.
This can be done in the following way: Start from one component and form all possible closed paths with the solid arrows starting and ending at the same component.
Then discard those loops that change the spin structure.   
The remaining paths form the desired complete set of remaining symmetries.
This leads to the following symmetry counts: $a^{P \sigma \sigma}$: $3$, $a^{P \uparrow \downarrow}$: $1$, $a^{X \uparrow \downarrow}$: $1$, $a^{D \sigma \sigma}$: $2$, $a^{D \uparrow \downarrow}$: $1$.

In order to classify these symmetries, we use the shortindex notation introduced in \eqref{definition_short_indices}, i.e. we encode the spatial structure in a (frequency dependent) block-matrix $A(\Omega) = \{A^{lk}_{ji} \}(\Omega)$, with a bosonic frequency $\Omega$.
To simplify notation, let us define the following generic independent transformations in position and frequency space: 
\begin{subequations}
\begin{align}
[\changed{A^{I_1}}]^{lk}_{ji}(\Omega) &= -A^{(-l)k}_{(j+l)i}(\Omega), \\
[\changed{A^{I_2}}]^{lk}_{ji}(\Omega) &= -A^{l(-k)}_{j(i+k)}(\Omega), \\
[\changed{A^T}]^{lk}_{ji}(\Omega) &= A^{kl}_{ij}(\Omega), \\
[\changed{A^Z}]^{lk}_{ji}(\Omega) &= A^{*(-l)(-k)}_{(j+l)(i+k)}(-\Omega).
\end{align} 
\label{spatial_symmetries}
\end{subequations}
With this, we can classify the symmetries in position and frequency as in Table~\ref{table_sym}. 
\begin{table}
\caption{Symmetries of vertex components in position and frequency space.}
\begin{center}
\begin{tabular}{ c|c|c|c|c|c }
       & $a^{P \sigma \sigma}$ & $a^{P \uparrow \downarrow}$ & $a^{X \uparrow \downarrow}$ & $a^{D \sigma \sigma}$ & $a^{D \uparrow \downarrow}$\\ 
\hline
$\changed{I_1}$  & $\checkmark$ & $-$          & $-$          & $-$          & $-$\\ 
$\changed{I_2}$  & $\checkmark$ & $-$          & $-$          & $-$          & $-$\\ 
$\changed{T}$    & $\checkmark$ & $\checkmark$ & $\checkmark$ & $\checkmark$ & $-$\\ 
$\changed{Z}$    & $-$          & $-$          & $-$          & $\checkmark$ & $\checkmark$ 
\label{table_sym}
\end{tabular}
\end{center}
\end{table}
The invariance under transposition $\changed{T}$ implies that for all vertex components in \eqref{spin_components} except $a^{D \uparrow \downarrow}$, the spatial block-matrix is symmetric, i.e. we only need to compute components with
\begin{align}
k\geq l, \label{symmetric_condition}
\end{align}
 and for $k=l$ it suffices to compute components with $i\geq j$ .
The additional symmetries $\changed{I_1}$, $\changed{I_2}$ in $a^{P \sigma \sigma}$ imply that there we only need to consider $l>0$. 
Finally, for both the $D$-channel contributions $a^{D \sigma \sigma}$ and $a^{D \uparrow \downarrow}$ we need to only compute the contributions for the frequencies $\Delta \geq 0$. 

\subsubsection*{Zero magnetic field}
In our work, we do not consider a finite magnetic field. This directly implies that we only need to compute one spin component of $a^{P \sigma \sigma}$ and $a^{D \sigma \sigma}$ (e.g. $\sigma = \uparrow$).
Furthermore, applying the same method as described above, we find that each of the mixed spin components now has one symmetry more, changing the symmetry counts to 
$a^{P \sigma \sigma}$: $3$, $a^{P \uparrow \downarrow}$: $2$, $a^{X \uparrow \downarrow}$: $2$, $a^{D \sigma \sigma}$: $2$, $a^{D \uparrow \downarrow}$: $2$.

Again we can classify the symmetries, see Table~\ref{table_sym_B_zero}.
\begin{table}
\caption{Same as in Table~\ref{table_sym} but for zero magnetic field.}
\begin{center}
\begin{tabular}{ c|c|c|c|c|c }
                 & $a^{P \uparrow \uparrow}$ & $a^{P \uparrow \downarrow}$ & $a^{X \uparrow \downarrow}$ & $a^{D \uparrow \uparrow}$ & $a^{D \uparrow \downarrow}$\\ 
\hline
$\changed{I_1}$            & $\checkmark$ & $-$          & $-$          & $-$          & $-$\\ 
$\changed{I_2}$            & $\checkmark$ & $-$          & $-$          & $-$          & $-$\\ 
$\changed{I \equiv I_1 \circ I_2}$  & $\checkmark$ & $\checkmark$ & $-$          & $-$          & $-$\\ 
$\changed{T}$              & $\checkmark$ & $\checkmark$ & $\checkmark$ & $\checkmark$ & $\checkmark$\\ 
$\changed{Z}$              & $-$          & $-$          & $\checkmark$ & $\checkmark$ & $\checkmark$ 
\label{table_sym_B_zero}
\end{tabular}
\end{center}
\end{table}
In terms of independent vertex components this implies that now we have to compute only the components with non-negative frequencies in the $X$-channel and that the spatial block structure of $a^{D \uparrow \downarrow}$ is now symmetric.
Furthermore, additionally to the symmetric condition \eqref{symmetric_condition}, now one only needs to compute the components with $l \leq 0$ in $a^{P \uparrow \downarrow}$. (Note that, in agreement with our choice of sign in \eqref{symmetric_condition}, this is a weaker statement than the condition $l > 0$ that is encountered for $a^{P \uparrow \uparrow}$, which is symmetric under $\changed{I_1}$ and $\changed{I_2}$ independently).    

\subsubsection*{Parity}
Finally, in the equilibrium context, the setup studied in this work is parity symmetric, due to the parity symmetry of the Hamiltonian. In our notation, the parity transformation can be expressed as 
\begin{align}
[\changed{A^M}]^{lk}_{ji}(\Omega) &= A^{(-l)(-k)}_{(-j)(-i)}(\Omega). \\
\label{par_sym}
\end{align}
In our work this relation is then a symmetry for all vertex components. 

\subsubsection*{Summary}
Each of the above-mentioned symmetries reduces the independent components of the vertex by roughly a factor of $1/2$.
Since in our work the computation of the bubbles \eqref{bubble_keldysh_components} takes the most time, our implementation does not make explicit use of the vertex symmetries in Table~\ref{table_sym_B_zero}.
However, they are useful tools for checking an implementation for possible mistakes.  

\changed{
\section{Derivation of vertex FDTs} \label{app_derivation_fdts}
In this section, we give a brief derivation of the vertex FDTs \eqref{vertex_fdts} and the relation \eqref{ad_connection}.
As starting point, we use the general statement \eqref{gamma_fdt} for the exact two-particle vertex in contour space, derived in \cite{Jakobs2009}, Eq.~(3.106).
%\begin{widetext}
%\begin{subequations}
%\begin{align}
%\Rep \gamma^{j'|j}_{\epsilon_1^{j'| j}}(\omega' | \omega) &= - \Big[1 - 2 f\Big(\Delta^{j'|j}(\omega'|\omega) + \mu\Big) \Big] \Rep \gamma^{j' | j}_{-\epsilon_1^{j'|j}}(\omega'| \omega), \\
%\Imp \gamma^{j'|j}_{-\epsilon_1^{j'| j}}(\omega' | \omega)&= - \Big[1 - 2 f\Big(\Delta^{j'|j}(\omega'|\omega) + \mu\Big) \Big] \Imp \gamma^{j' | j}_{\epsilon_1^{j'|j}}(\omega'| \omega),
%\end{align}
%\label{gamma_fdt}
%\end{subequations}
%\end{widetext}
%where for given $\epsilon = \pm 1$
%\begin{align}
%\gamma^{j'| j}_\epsilon = \sum_{\stackrel{\alpha',\alpha}{(-1)^{\sum_k (\alpha_k' + \alpha_k)}=\epsilon}} D^{j' | \alpha'} \gamma^{\alpha'| \alpha} (D^{-1})^{\alpha | j}, 
%\label{gamma_epsilon}
%\end{align}
%with the Keldysh rotation
%\begin{subequations}
%\begin{align}
%D^{-|1} &= D^{\pm | 2} = \frac{1}{\sqrt{2}}, \\ 
%D^{+|1} &= - \frac{1}{\sqrt{2}}.
%\end{align}
%\end{subequations}
%Furthermore,
%\begin{align}
%\epsilon_1^{j'| j} = (-1)^{m^{j'| j}}, \\ 
%m^{j'| j} = \sum_{k: j_k = +} 1 - \sum_{k: j_k'=+} 1,
%\end{align}
%and 
%\begin{align}
%\Delta^{j'| j}(\omega' | \omega) = \sum_{k: j_k=+} (\omega_k - \mu) - \sum_{k:j_k'=+}(\omega_k' - \mu).
%\end{align}
We remark that the spin and spatial structure of \eqref{gamma_fdt} is trivial. 
For this reason, we will not display any spin or spatial indices in this section.   

Inserting the channel decomposition \eqref{channel_decomposition} in \eqref{gamma_fdt} yields
\begin{widetext}
\begin{subequations}
\begin{align}
\Rep \Big[\nu^{j'|j}_{\epsilon_1^{j'|j}} + \sum_A (\varphi^A_{\epsilon_1^{j'|j}})^{j'|j}(\Omega^A) \Big] &=-\Big[1 - 2f\Big(\Delta^{j'|j}(\Pi,\Chi,\Delta) + \mu\Big) \Big] \Rep \Big[\nu^{j'|j}_{-\epsilon_1^{j'|j}} + \sum_A (\varphi^A_{-\epsilon_1^{j'|j}})^{j'|j}(\Omega^A) \Big], \label{phi_fdt_real} \\
\Imp \Big[\nu^{j'|j}_{-\epsilon_1^{j'|j}} + \sum_A (\varphi^A_{-\epsilon_1^{j'|j}})^{j'|j}(\Omega^A) \Big] &=-\Big[1 - 2f\Big(\Delta^{j'|j}(\Pi,\Chi,\Delta) + \mu\Big) \Big] \Imp \Big[\nu^{j'|j}_{\epsilon_1^{j'|j}} + \sum_A (\varphi^A_{\epsilon_1^{j'|j}})^{j'|j}(\Omega^A) \Big], \label{phi_fdt_imag}
\end{align}
\label{phi_fdt}
\end{subequations}
\end{widetext}
where $A \in \{P,X,D \}$ and correspondingly $\Omega^A \in \{\Pi,\Chi,\Delta \}$, and where we applied an analogous definition of \eqref{gamma_epsilon} to the $\varphi$'s and $\nu$. 
Using \eqref{bosonic_frequencies}, we obtain for $\Delta^{j'| j}(\omega' | \omega)$ 
\begin{subequations}
\begin{align}
\Delta^{-- | --}(\omega' | \omega) &= 0, \\
\Delta^{++ | --}(\omega' | \omega) &= 2\mu -(\omega_1' + \omega_2') = 2\mu - \Pi, \\
\Delta^{-+ | +-}(\omega' | \omega) &= \omega_1 - \omega_2' = -X, \\
\Delta^{-+ | -+}(\omega' | \omega) &= \omega_2 - \omega_2' = \Delta.
\end{align} 
\label{delta_relations}
\end{subequations}
%If we express $\gamma$ in \eqref{gamma_epsilon} in terms of the channel decomposition \eqref{channel_decomposition} and use \eqref{delta_relations} , 
%we obtain i.a.\ to the following relations 
%\begin{widetext}
%\begin{align}
%\Rep \Big[\nu^{--|--}_+ + \sum_A (\varphi^A_+)^{--|--}(\Omega^A) \Big] &= 0, \\
%\Rep \Big[\nu^{++|--}_+ + \sum_A (\varphi^A_+)^{++|--}(\Omega^A) \Big] &=-\Big[1 - 2f(-\Pi + \mu) \Big] \Rep \Big[\nu^{++|--}_- + \sum_A (\varphi^A_-)^{++|--}(\Omega^A) \Big], \\
%\Rep \Big[\nu^{-+|+-}_+ + \sum_A (\varphi^A_+)^{-+|+-}(\Omega^A) \Big] &=-\Big[1 - 2f(-\Chi + \mu) \Big] \Rep \Big[\nu^{-+|+-}_- + \sum_A (\varphi^A_-)^{-+|+-}(\Omega^A) \Big], \\
%\Rep \Big[\nu^{-+|-+}_+ + \sum_A (\varphi^A_+)^{-+|-+}(\Omega^A) \Big] &=-\Big[1 - 2f(\Delta + \mu) \Big] \Rep \Big[\nu^{-+|-+}_- + \sum_A (\varphi^A_-)^{-+|-+}(\Omega^A) \Big],  
%\label{fdt_cont}
%\end{align}
%\end{widetext}
%where $A \in \{P,X,D \}$ and correspondingly $\Omega^A \in \{\Pi,\Chi,\Delta \}$, and where we applied an analogous definition of \eqref{gamma_epsilon} to the $\varphi$'s and $\nu$. 
Furthermore, combining \eqref{gamma_epsilon} and \eqref{barevertex_keldysh_structure} yields the bare vertex expressions 
\begin{subequations}
\begin{align}
\nu^{j'| j}_+&=0, \\ 
\nu^{j_1' j_2' | j_1 j_2}_- &= \nu^{j_1' j_2' | j_1 j_2} \sim \delta(j_1' = j_2' = j_1 = j_2).     
\end{align}
\label{form_barevertex_contour}
\end{subequations}
Analogously, a combination of \eqref{gamma_epsilon} with the Keldysh structure of the vertices \eqref{keldysh_components} leads i.a.\ to the relations 
\begin{subequations}
\begin{align}
(\varphi^A_-)^{--|--} &= a^A + d^A, \\
(\varphi^A_+)^{--|--} &= b^A,
\end{align}
for all $A \in \{P,X,D \}$, as well as
\begin{align}
(\varphi^P_-)^{++ | --}&=-a^P+d^P, \\
(\varphi^P_+)^{++ | --}&=-b^P, \\ 
(\varphi^P_{\pm})^{-+|+-} &=(\varphi^P_{\pm})^{-+|-+} =  0, 
\end{align}

\begin{align}
(\varphi^X_-)^{-+ | +-}&=a^X-d^X, \\
(\varphi^X_+)^{-+ | +-}&=-b^X, \\ 
(\varphi^X_\pm)^{++ | --}&= (\varphi^X_\pm)^{-+ | -+}=0, 
\end{align}
and
\begin{align}
(\varphi^D_-)^{-+ | -+}&=a^D-d^D, \\
(\varphi^D_+)^{-+ | -+}&=-b^D, \\
(\varphi^D_\pm)^{++ | --}&= (\varphi^D_\pm)^{-+ | +-} = 0. 
\end{align}
\label{phi_relations}
\end{subequations}

If we insert \eqref{delta_relations}, \eqref{form_barevertex_contour} and \eqref{phi_relations} into \eqref{phi_fdt_real}, we obtain   
\begin{subequations}
\begin{align}
\sum_A \Rep\Big[b^A&(\Omega^A)\Big] = 0, \\
\Rep\Big[-b^P(\Pi)\Big] &= -\Big[1 - 2f(3\mu - \Pi) \Big] \Rep\Big[-a^P + d^P\Big](\Pi), \\
\Rep\Big[-b^X(\Chi)\Big] &= -\Big[1 - 2f(\mu - \Chi) \Big] \Rep\Big[a^X - d^X\Big](\Chi), \\
\Rep\Big[-b^D(\Delta)\Big] &= -\Big[1 - 2f(\mu + \Delta) \Big] \Rep\Big[a^D - d^D\Big](\Delta). 
\end{align}
\label{fdt_ad_real}
\end{subequations}
If we insert \eqref{delta_relations}, \eqref{form_barevertex_contour} and \eqref{phi_relations} into \eqref{phi_fdt_imag}, we obtain
\begin{subequations}
\begin{align}
\sum_A \Imp\Big[ a^A(\Omega^A)& + d^A(\Omega^A) \Big] = 0, \label{fdt_ad_imag_1}\\
\Imp\Big[-a^P + d^P\Big](\Pi) &= -\Big[1 - 2f(3\mu - \Pi) \Big] \Imp\Big[-b^P(\Pi) \Big], \label{fdt_ad_imag_2} \\
\Imp\Big[a^X - d^X\Big](\Chi) &= -\Big[1 - 2f(\mu - \Chi) \Big] \Imp\Big[-b^X(\Chi) \Big], \\
\Imp\Big[a^D - d^D\Big](\Delta) &= -\Big[1 - 2f(\mu + \Delta) \Big] \Imp\Big[- b^D\Big](\Delta). \label{fdt_ad_imag_4} 
\end{align}
\label{fdt_ad_imag}
\end{subequations}
Using \eqref{fdt_ad_real} and \eqref{fdt_ad_imag_1} together with the continuity of the vertex components as well as their high frequency asymptotic $\lim_{|\Omega| \rightarrow \infty} \varphi^A(\Omega) = 0 $ yields relation \eqref{ad_connection}.
If we additionally also use the relations (\ref{fdt_ad_imag_2}-\ref{fdt_ad_imag_4}) and the identity 
\begin{align}
\frac{1}{1-2f(\mu + \Omega)} = \coth \Big(\frac{\Omega}{2T} \Big), 
\end{align}
we obtain the vertex FDTs \eqref{vertex_fdts}.
}

\changed{
\section{Explicit flow equations} \label{app_explicit_flow}
In this section, we give the full form of the flow equations discussed in Sec.~\ref{sec_flow_equations}, including all spin- and spatial indices. 
For the notation of the latter, we use the general short-index notation introduced in \eqref{definition_short_indices}.} 
Using the symmetries of the vertex for the equilibrium case (as discussed in App.~\ref{app_sym}) , the general fRG-flow equations in the channel decomposition (see e.g. \cite{Jakobs2009,Jakobs2010a}) can be formulated as shown below. 

\changed{
In order to facilitate the representation of the self-energy flow, it is convenient to split the self-energy into a static and a dynamic contribution $\Sigma = \Sigma_s + \Sigma_d$. 
Furthermore, we introduce first the following auxiliary quantities, identified by a tilde:}
\begin{widetext}
\changed{
\begin{subequations}
\begin{align}
    \partial_\Lambda (\tilde\Sigma_s)^{R \sigma}_{j (j+l)}(\omega) 
&=  -\frac{i}{2\pi} \int d\omega' \,
       \Big[\frac{1}{2} \changed{\bar{v}}^{\sigma \sigma | \sigma \sigma}_{j (i+k) | (j+l) i} + (a^D)^{\sigma \sigma l k}_{j i}(0) \Big] 
           S^{K \sigma}_{i (i+k)}(\omega'), \label{flow_self_sigma_static} 
\end{align}
\begin{align}
\partial_\Lambda (\tilde\Sigma_d)^{R \sigma}_{j i}(\omega) 
=\frac{i}{2\pi} \int d\omega' \, \Big\{
    &(b^D)^{\sigma \sigma l k}_{j i }(\omega-\omega') S^{R \sigma}_{(j+l) (i+k)}(\omega') 
    -(b^P)^{\sigma \sigma l k}_{j i }(\omega'+\omega) S^{A \sigma}_{(i+k) (j+l)}(\omega') \nonumber \\
   +&\Big[(a^D)^{\sigma \sigma l k}_{j i }(\omega-\omega') -(a^P)^{\sigma \sigma l k}_{j i }(\omega'+\omega) \Big] S^{K \sigma}_{(j+l)(i+k)}(\omega')  \Big\}. \label{flow_self_sigma_dynamic}
\end{align}
\label{flow_self_sigma}
\end{subequations}

Then the flow of the self-energy is given by:
\begin{subequations}
\begin{align}
   \partial_\Lambda (\Sigma_s)^{R \uparrow}_{j (j+l) }(\omega)  
&= \partial_\Lambda (\tilde\Sigma_s)^{R \uparrow}_{j (j+l) }(\omega) \nonumber\\
& - \frac{i}{2\pi} \int d\omega' \,
       \Big[ \frac{1}{2} \changed{\bar{v}}^{\uparrow \downarrow | \uparrow \downarrow}_{j (i+k) | (j+l) i} +(a^D)^{\uparrow \downarrow l k}_{j i}(0) \Big] 
           S^{K \downarrow}_{i (i+k)}(\omega'), \label{flow_self_up}
\end{align}
\begin{align}
   \partial_\Lambda (\Sigma_d)^{R \uparrow}_{j i}(\omega)  
&= \partial_\Lambda (\tilde\Sigma_d)^{R \uparrow}_{j i }(\omega) \nonumber\\
&- \frac{i}{2\pi} \int d\omega' \,
     \Big\{ 
          (b^X)^{\uparrow \downarrow  l k}_{j i }(\omega'-\omega)
           S^{R \downarrow}_{(j+l) (i+k)}(\omega') 
         +  (b^P)^{\uparrow \downarrow l k }_{j i }(\omega'+\omega) 
           S^{A \downarrow}_{(i+k) (j+l)}(\omega') \nonumber \\ 
       & + \Big[(a^X)^{\uparrow \downarrow l k}_{j i }(\omega'-\omega) + (a^P)^{\uparrow \downarrow l k}_{j i }(\omega'+\omega)  \Big]
           S^{K \downarrow}_{(j+l)(i+k)}(\omega') \Big\}, 
\end{align}
\label{self_flow_up}
\end{subequations}
and
\begin{subequations}
\begin{align}
   \partial_\Lambda (\Sigma_s)^{R \downarrow}_{j (j+l)}(\omega) 
&= \partial_\Lambda (\tilde\Sigma_s)^{R \downarrow}_{j (j+l) }(\omega) \nonumber\\
& - \frac{i}{2\pi} \int d\omega' \,
       \Big[ \frac{1}{2} \changed{\bar{v}}^{\uparrow \downarrow | \uparrow \downarrow}_{i  (j+l)| (i+k) j } + (a^D)^{\uparrow \downarrow k l }_{i j }(0) \Big] 
           S^{K \uparrow}_{i (i+k)}(\omega') . \label{flow_self_down}
\end{align}
\begin{align}
   \partial_\Lambda (\Sigma_d)^{R \downarrow}_{j i}(\omega) 
&= \partial_\Lambda (\tilde\Sigma_d)^{R \downarrow}_{j i}(\omega) \nonumber\\
& - \frac{i}{2\pi} \int d\omega' \,
     \Big\{ (b^X)^{\uparrow \downarrow (-l)(-k) }_{(j+l)(i+k)}(\omega - \omega') S^{R \uparrow}_{(j+l)(i+k)}(\omega')
         + (b^P)^{\uparrow \downarrow (-l) (-k)}_{(j+l)(i+k)}(\omega'+\omega) S^{A \uparrow}_{(i+k)(j+l)}(\omega') \nonumber \\
       & + \Big[ (a^{X*})^{\uparrow \downarrow (-l) (-k)}_{(j+l) (i+k) }(\omega-\omega')   
         + (a^P)^{\uparrow \downarrow (-l) (-k)}_{(j+l) (i+k) }(\omega'+\omega) \Big] S^{K \uparrow}_{(j+l) (i+k)}(\omega') \Big\}. 
\end{align}
\label{self_flow_down}
\end{subequations}

Before we proceed to write down the flow of the two-particle vertex, let us take a look at the bubble terms \eqref{bubbles_elementar}. 
Displaying the full spin and spatial structure, \eqref{bubbles_elementar} reads
\begin{subequations}
\begin{align}
\Big[(\tilde{I}^{pp})^{\alpha_1' \alpha_2' | \alpha_1 \alpha_2}\Big]^{ \sigma \tau lk}_{ji}(\Pi) &= \frac{i}{2\pi} \int d\omega \Big[ (S^{\alpha_1' \alpha_1})^\sigma_{ji}(\omega) (G^{\alpha_2' \alpha_2})^\tau_{(j+l)(i+k)}(\Pi - \omega) + [S \leftrightarrow G] \Big], \label{bubbles_elementar_indices_pp}\\
\Big[(\tilde{I}^{ph})^{ \alpha_1' \alpha_2' | \alpha_1 \alpha_2}\Big]^{\sigma \tau lk}_{ji}(X) &= \frac{i}{2\pi} \int d\omega \Big[ (S^{\alpha_1' \alpha_1})^\sigma_{ji}(\omega) (G^{\alpha_2' \alpha_2})^\tau_{(i+k)(j+l)}(\omega + X) + [S \leftrightarrow G] \Big].\label{bubbles_elementar_indices_ph}
\end{align}
\label{bubbles_elementar_indices}
\end{subequations}
The symmetrical appearance of $G$ and $S$ in definition \eqref{bubbles_elementar_indices} implies a corresponding symmetry for the whole bubbles.  
%In order to facilitate notation, we use in the following abbreviations for a general block-matrix quantity $A^{lk}_{ji}$
%\begin{subequations}
%\begin{align}
%(A^T)^{lk}_{ji} &= [t(A)]^{lk}_{ji} = A^{kl}_{ij}, \\
%(A^I)^{lk}_{ji} &= [(i_1 \circ i_2)(A)]^{lk}_{ji} = A^{(-l)(-k)}_{(j+l)(i+k)}. 
%\end{align} 
%\end{subequations}
Using the notation introduced in \eqref{spatial_symmetries} with $I \equiv I_1 \circ I_2$, the implied $[G \leftrightarrow S]$ symmetry of the bubble reads 
\begin{subequations}
\begin{align}
\Big[(\tilde{I}^{pp})^{\alpha_1' \alpha_2' | \alpha_1 \alpha_2}\Big]^{\sigma \tau}(\Pi) &= \Big[(\tilde{I}^{pp})^{\alpha_2' \alpha_1' | \alpha_2 \alpha_1}\Big]^{I \tau \sigma}(\Pi), \label{bubble_GS_sym_pp}\\
\Big[(\tilde{I}^{ph})^{\alpha_1' \alpha_2' | \alpha_1 \alpha_2}\Big]^{\sigma \tau}(X)   &= \Big[(\tilde{I}^{ph})^{\alpha_2' \alpha_1' | \alpha_2 \alpha_1}\Big]^{I \tau \sigma}(-X).\label{bubble_GS_sym_ph}
\end{align}
\label{bubble_GS_sym}
\end{subequations}
These symmetries immediately follow from definiton \eqref{bubbles_elementar_indices}.
Additionally, by complex conjugation, we have for $\zeta \in \{I^{pp},I^{ph} \}$
\begin{align}
\zeta^{\alpha_1' \alpha_2' | \alpha_1 \alpha_2} = (-1)^{1 + \alpha_1' + \alpha_2' + \alpha_1 + \alpha_2} \Big[\zeta^{\alpha_1 \alpha_2 | \alpha_1' \alpha_2'} \Big]^{*},
\label{bubble_complex_conjugation}
\end{align}
which follows from \eqref{bubbles_elementar_indices} and \eqref{propagators_hermitian_conjugation}.
In terms of the components in Keldysh space \eqref{bubble_keldysh_components}, and with properly treated spin and spatial structure, the bubbles $I^{A}$ with $A \in \{P,X,D \}$ from \eqref{channel_bubbles} take the form 
\begin{subequations}
\begin{align}
(I^P)^{\sigma \tau}(\Pi) &= \Big[(\tilde{I}^{pp})^{22|21} + (\tilde{I}^{pp})^{22|12}\Big]^{\sigma \tau}(\Pi) = \Big[(I^{pp})^{ \sigma \tau } + (I^{pp})^{I \tau \sigma }\Big](\Pi), \label{bubbles_spin_spatial_P}\\
(I^{X})^{\sigma \tau}(\Chi) &= \Big[ (\tilde{I}^{ph})^{22|12} + (\tilde{I}^{ph})^{21|22} \Big]^{\sigma \tau}(\Chi) =  \Big[ (I^{ph})^{ \sigma \tau }(\Chi) + (I^{ph})^{I* \tau \sigma}(-\Chi) \Big], \label{bubbles_spin_spatial_X}\\
(I^{D})^{\sigma \tau}(\Delta)&=- \Big[ (\tilde{I}^{ph})^{22|21} + (\tilde{I}^{ph})^{12|22} \Big]^{I \sigma \tau}(\Delta) =  -(I^X)^{\tau \sigma}(-\Delta) \Big]. \label{bubbles_spin_spatial_D}
\end{align}
\label{bubbles_spin_spatial}
\end{subequations}
Furthermore, using the propagator FDTs \eqref{propagator_fdt}, together with the general relation 
\begin{align}
1 - 2 f(\mu - \omega) = -\Big[1 - 2 f(\mu + \omega)\Big],
\label{anti_sym_kel_prefactor}
\end{align}
one can straightforwardly show (c.f.\ \cite{Jakobs2009}, pp.~166-167) that the bubbles \eqref{bubbles_spin_spatial} are real at their feedback frequencies, i.e.\ $I^P(2\mu)$ and $I^X(0)$, $I^D(0)$ are real.
}

For the flow of the vertex we define:
\begin{subequations}
\begin{align} 
(\changed{\tilde{a}^P})^{\sigma \sigma l k}_{ji}(\Pi)          &= \frac{1}{2} \changed{\bar{v}}^{\sigma \sigma | \sigma \sigma}_{j(j+l)|i(i+k)}
                                              + (a^P)^{\sigma \sigma lk}_{ji}(\Pi) 
                                              \changed{- (\phi^D})^{\sigma \sigma (i+k-j)(j+l-i)}_{j i} 
                                              + (\phi^D)^{\sigma \sigma (i-j)(j+l-i-k)}_{j(i+k)}, \label{tilde_a_p} \\
(\changed{\tilde{a}^P})^{\uparrow \downarrow l k}_{ji}(\Pi)    &= \frac{1}{2} \changed{\bar{v}}^{\uparrow \downarrow | \uparrow \downarrow}_{j(j+l)|i(i+k)}
                                             + (a^P)^{\uparrow \downarrow lk}_{ji}(\Pi) 
                                             + (\phi^X)^{\uparrow \downarrow (i+k-j)(j+l-i)}_{ji} 
                                             + (\phi^D)^{\uparrow \downarrow (i-j)(j+l-i-k)}_{j(i+k)}, \\
(\changed{\tilde{a}^X})^{\uparrow \downarrow l k}_{ji}(\Chi)   &= \frac{1}{2} \changed{\bar{v}}^{\uparrow \downarrow | \uparrow \downarrow}_{j(i+k)|i(j+l)}
                                             + (a^X)^{\uparrow \downarrow lk}_{ji}(\Chi) 
                                             + (\phi^{\changed{P}})^{\uparrow \downarrow (i+k-j)(j+l-i)}_{ji} 
                                             + (\phi^D)^{\uparrow \downarrow (i-j)\changed{(i+k-j-l)}}_{j\changed{(j+l)}}, \label{tilde_a_x} \\
(\changed{\tilde{a}^D})^{\sigma \sigma l k}_{ji}(\Delta)       &= \frac{1}{2} \changed{\bar{v}}^{\sigma \sigma | \sigma \sigma}_{j(i+k)|(j+l)i}
                                             + (a^D)^{\sigma \sigma lk}_{ji}(\Delta) 
                                             + (\phi^P)^{\sigma \sigma (i+k-j)(i-j-l)}_{j(j+l)} 
                                             - (\phi^D)^{\sigma \sigma (i-j)(i+k-j-l)}_{j(j+l)}, \\
(\changed{\tilde{a}^D})^{\uparrow \downarrow l k}_{ji}(\Delta) &= \frac{1}{2} \changed{\bar{v}}^{\uparrow \downarrow | \uparrow \downarrow}_{j(i+k)|(j+l)i}
                                              + (a^D)^{\uparrow \downarrow lk}_{ji}(\Delta) 
                                              + (\phi^P)^{\uparrow \downarrow (i+k-j)(i-j-l)}_{j(j+l)} 
                                              + (\phi^X)^{\uparrow \downarrow (i-j)(i+k-j-l)}_{j(j+l)}.
\end{align} 
\label{vertex_rhs}
\end{subequations}
The static interchannel feedback is chosen as in \cite{Jakobs2009,Jakobs2010a,Schimmel2017} 
$\phi^P = a^P(2\mu)$,
$\phi^X = a^X(0)$,
$\phi^D = a^D(0)$.
\changed{
Note that since the bubbles \eqref{bubbles_spin_spatial} are real valued at the respective feedback frequencies, the $\phi$ are also real and furthermore (due to the vertex FDTs \eqref{vertex_fdts} and \eqref{ad_connection}) they have the same Keldysh structure as the bare vertex \eqref{barevertex_keldysh_structure}.  } 
\end{widetext}
\changed{
If we use the definition of block-matrix multiplication in spacial indices \eqref{block_matrix_multiplication},}
%\begin{align}
%[A B]^{lk}_{ji}=A^{l k_1}_{j i_1} B^{k_1 k}_{i_1 i}, 
%\end{align}
%and define a transposition in spacial indices
%\begin{align}
%[A^T]^{lk}_{ji} = A^{kl}_{ij},
%\end{align}
the flow of the vertex can be written in the simple form:
\begin{subequations}
\begin{align}
(\dot{a}^P)^{\sigma \sigma}(\Pi) 
&= \frac{1}{2}(\changed{\tilde{a}^P})^{\sigma \sigma}(\Pi) \cdot 
              (I^P)^{\sigma \sigma}(\Pi) \cdot
              (\changed{\tilde{a}^P})^{\sigma \sigma}(\Pi) \label{flow_pss} \\
(\dot{a}^P)^{\uparrow \downarrow}(\Pi)  
&= (\changed{\tilde{a}^P})^{\uparrow \downarrow}(\Pi)\cdot 
   (I^P)^{\uparrow \downarrow}(\Pi)\cdot
   (\changed{\tilde{a}^P})^{\uparrow \downarrow}(\Pi) \\
(\dot{a}^X)^{\uparrow \downarrow}(\Chi) 
&=(\changed{\tilde{a}^X})^{\uparrow \downarrow}(\Chi)\cdot
  (I^X)^{\uparrow \downarrow}(\Chi)\cdot
  (\changed{\tilde{a}^X})^{\uparrow \downarrow}(\Chi) \\
(\dot{a}^D)^{\uparrow \uparrow}(\Delta)
&=- (\changed{\tilde{a}^D})^{\uparrow \uparrow}(\Delta)\cdot
    (I^X)^{\uparrow \uparrow}(-\Delta)\cdot
   (\changed{\tilde{a}^D})^{\uparrow \uparrow}(\Delta) \nonumber \\
&\phantom{=\ }- (\changed{\tilde{a}^D})^{\uparrow \downarrow}(\Delta)\cdot
    (I^X)^{\downarrow \downarrow}(-\Delta)\cdot
   (\changed{\tilde{a}^D})^{T\uparrow \downarrow}(\Delta) \\
(\dot{a}^D)^{\downarrow \downarrow}(\Delta)
&=- (\changed{\tilde{a}^D})^{\downarrow \downarrow}(\Delta)\cdot
    (I^X)^{\downarrow \downarrow}(-\Delta)\cdot
   (\changed{\tilde{a}^D})^{\downarrow \downarrow}(\Delta) \nonumber \\
&\phantom{=\ }- (\changed{\tilde{a}^D})^{T\uparrow \downarrow}(\Delta)\cdot
    (I^X)^{\uparrow \uparrow}(-\Delta)\cdot
   (\changed{\tilde{a}^D})^{\uparrow \downarrow}(\Delta) \\
(\dot{a}^D)^{\uparrow \downarrow}(\Delta) 
&= -(\changed{\tilde{a}^D})^{\uparrow \downarrow}(\Delta)\cdot
    (I^X)^{\downarrow \downarrow}(-\Delta)\cdot
   (\changed{\tilde{a}^D})^{\downarrow \downarrow}(\Delta) \nonumber \\
&\phantom{=\ } -(\changed{\tilde{a}^D})^{\uparrow \uparrow}\cdot
    (I^X)^{\uparrow \uparrow}(-\Delta)\cdot
   (\changed{\tilde{a}^D})^{\uparrow \downarrow}. \label{flow_dud} 
\end{align}
\label{flow_complete}
\end{subequations}

%Furthermore, the initial conditions for our fRG flow from a finite but large $\Lambda_{\text{ini}}$ (in practice $\Lambda_{\text{ini}}=10^5\tau$) are given by \cite{Jakobs2009,Jakobs2010a} 
%\begin{align}
%\Sigma^{R \sigma \Lambda_{\text{ini}} }_{ij}(\omega) &= \frac{1}{2} \sum_{k\tau} \changed{\bar{v}}^{\sigma \tau | \sigma \tau}_{i k | j k}, \\
%a^{P \Lambda_{\text{ini}}} &= a^{X \Lambda_{\text{ini}}} = a^{D \Lambda_{\text{ini}}} = 0.
%\end{align}

\section{Importance of feedback frequencies}
\label{importance_feedback}
In this section, we discuss the importance of the feedback frequencies in the vertex (c.f.\ Sec.~\ref{dynamic_feedback_length}) for low-energy observables.
In particular, we use the linear response conductance $g$ of Eq.~\eqref{conductance_1} as an example. 
In order to illustrate the underlying mechanism, we first focus on the system at $T=0$.  
In this case, the conductance consists only of the one-particle contribution \eqref{op_g}, i.e.\ it is completely determined by the knowledge of $\Sigma(\mu)$. 
We obtain $\Sigma(\mu)$ via our fRG flow, i.e.\ in order to understand the influence of our treatment of the two-particle vertex on the conductance, we have to take a look at the flow equations formulated in Sec.~\ref{sec_flow_equations}. 
\changed{
In case of the static part $\partial_\Lambda(\Sigma_s)^R$, this is easy: The vertex contribution $a^D$ is only evaluated directly at the feedback frequency $\Delta = 0$. 
For the dynamic contribution $\partial_\Lambda (\Sigma_d)^R$, we have to look a little closer.   
}
In the $T=0$ case, we can prove here two exact statements (\ref{imp_self_up},\ref{imp_self_down}).
By using the FDTS \eqref{vertex_fdts} and performing the limit $T\rightarrow 0$, we obtain 
\begin{widetext}
\changed{
\begin{align}
   \partial_\Lambda (\tilde\Sigma_d)^{R \sigma}_{j i}(\mu)  
&=  \frac{1}{\pi} \int d\omega' \, \Big( 2\theta(\omega' - \mu) -1 \Big)  \Imp \Big[ \Big( (a^P)^{\sigma \sigma l k}_{ji}(\mu + \omega') - (a^{D*})^{\sigma \sigma l k}_{ji}(\mu - \omega') \Big) S^{R \sigma}_{(j+l)(i+k)}(\omega') \Big].
\label{Sigma_tilde_at_mu_T0}
\end{align}
}
Since both $a^P$ and $a^D$ are retarded and approach constants and $S^R(\omega) \sim \frac{1}{\omega^2}$ for large frequency arguments $\omega$, we have furthermore: 
\begin{align}
\int d\omega' \Big( (a^P)^{\sigma \sigma l k}_{ji}(\mu + \omega') - (a^{D*})^{\sigma \sigma l k}_{ji}(\mu - \omega') \Big) S^{R \sigma}_{\changed{(j+l)(i+k)}}(\omega') = 0.
\end{align}
With this, we can rewrite \eqref{Sigma_tilde_at_mu_T0} and obtain
\changed{
\begin{align}
   \partial_\Lambda (\tilde\Sigma_d)^{R \sigma}_{j i }(\mu) 
&=  -\frac{2}{\pi} \int^\mu_{-\infty} d\omega' \, \Imp \Big[ \Big( (a^P)^{\sigma \sigma l k}_{ji}(\mu + \omega') - (a^{D*})^{\sigma \sigma l k}_{ji}(\mu - \omega') \Big) S^{R \sigma}_{(j+l)(i+k)}(\omega') \Big].
\label{imp_self_tilde}
\end{align}
}
Proceeding analogously, we can obtain for the complete \changed{dynamic} self-energy
\changed{
\begin{subequations}
\begin{align}
\partial_\Lambda (\Sigma_d)^{R \uparrow}_{j i }(\mu) &= \partial_\Lambda (\tilde\Sigma_d)^{R \uparrow}_{j | i }(\mu) 
                                                     - \frac{2}{\pi} \int^\mu_{-\infty} d\omega' \, \Imp \Big[ \Big\{ (a^P)^{\uparrow \downarrow lk}_{ji}(\mu + \omega')
                                                                                                                      +(a^{X*})^{\uparrow \downarrow lk}_{ji}(\omega'-\mu) \Big\} S^{R \downarrow}_{(j+l)(i+k)}(\omega') \Big], \label{imp_self_up} \\
\partial_\Lambda (\Sigma_d)^{R \downarrow}_{j i }(\mu) &= \partial_\Lambda (\tilde\Sigma_d)^{R \downarrow}_{j i }(\mu) 
                                                     - \frac{2}{\pi} \int^\mu_{-\infty} d\omega' \, \Imp \Big[ \Big\{ (a^P)^{\uparrow \downarrow (-l)(-k)}_{(j+l)(i+k)}(\mu + \omega')
                                                                                                                      +(a^{X})^{\uparrow \downarrow (-l)(-k)}_{(j+l)(i+k)}(\mu - \omega') \Big\} S^{R \uparrow}_{(j+l)(i+k)}(\omega') \Big]. \label{imp_self_down}
\end{align}
\label{imp_self}
\end{subequations}
}

\end{widetext}

In the one-particle part of the conductance \eqref{op_g}, we have to evaluate \changed{$G^{R\sigma}_{-N N}(\mu)$} at %\scrap{the} 
opposite ends of the chain. %\scrap{$i=-N$, $j=N$}.
\changed{
In order for a self-energy component $(\Sigma_d)^R_{ji}(\mu)$ to yield a substantial contribution to this propagator, the spatial indices $j,i$ have to fulfill at least one of the following two criteria: 
(i) The spatial indices lie on different sides of the QPC barrier. 
In this case, $\Sigma^R_{ji}(\mu)$ yields a direct hopping contribution to $G^{R\sigma}_{-N N}(\mu)$.
(ii) At least one spatial index lies in the region of the barrier top.
In this case, one either obtains a still significant hopping contribution (if the other index does not lie in the region of the barrier top) or a renormalization of the barrier top (if both indices lie in the region of the barrier top).
The remaining case, where both indices lie away from the barrier top on the same side of the QPC barrier, does not yield any significant contributions to the conductance. In this case, both spatial indices $j,i$ lie in a connected spatial region where the lower band edge is way below the chemical potential (c.f.\ Fig.~\ref{fig_model}(a)), i.e.\ in this region the movement of electrons is not impaired anyway.  
Therefore, we will assume in the following that $j,i$ fulfill at least one of the two criteria (i),(ii). 
}

%\changed{
%In this case, we can use that \changed{the interaction range of the bare interaction $L_U$ is finite and much shorter than the length of the entire system $2N=|i-j|$ (including the QPC flanks).}
%Therefore, we can approximately neglect the static terms $\partial_\Lambda(\Sigma_s)^{R\sigma}_{-N N}$ (\ref{flow_self_up},\ref{flow_self_down}), since the bare vertex $\bar{v}$ vanishes exactly and the $a^D(0)$ terms appear with an huge short index $|l|=2N$.
%Furthermore, in the dynamic terms \eqref{imp_self}, 
%}
In this case, we can approximately change the lower bound of the integration in (\ref{Sigma_tilde_at_mu_T0}-\ref{imp_self_down}) from $-\infty$ to $\omega_b=-2\tau + V_g$, the energy of the barrier top in the middle of the QPC:
For small $l,k$ the propagator $S^R_{(i+k)(j+l)}(\omega')$ gets suppressed exponentially by the barrier once $\omega' < \omega_b$. 
For large $l$ or $k$, the vertex contributions $(a^A)^{lk}(\omega')$ will be small, since the interaction range of the bare interaction is finite and much shorter than the length of the entire system (including the QPC flanks).     
Therefore, in the flow of the self-energy compontents $\Sigma^R_{j,i}(\mu)$ where $i,j$ fulfill at least one of the conditions (i) or (ii), only vertex components within the frequency range $[2\mu-(\mu-\omega_b),2\mu]$ are important for the P-contribution, and in the range $[-(\mu-\omega_b),(\mu-\omega_b)]$ for the X- and D-channel contributions. 
Since we are especially interested in the behavior during the first conductance step, i.e.\ when $(\mu-\omega_b) \sim \Omega_x$, the leading frequency contribution of the vertex components lies in the frequency range $\Omega^f \pm \Omega_x$, where $\Omega^f$ are the feedback frequencies \changed{$2\mu$ and $0$,} defined in Sec.~\ref{subsection_frequency}.
% \scrap{Generally speaking, these are the vertex components that contribute to the self-energy of interest in second-order of the bare interaction. 
% Checking the general flow equations \changed{\eqref{self_flow_up}, \eqref{self_flow_down} and \eqref{vertex_rhs} }, one can immediately see that contributions of the other vertex components are at least of fourth order or higher in the bare interaction}. 

At finite temperatures, for the one-particle contribution of the conductance, the same argument holds in essence.
It is just slightly more technical due to keeping track of the temperature smearing of Fermi steps. 
Instead of evaluating $\Sigma$ only at $\mu$, we now need it in an interval $[\mu-\Delta_T,\mu+\Delta_T$, where the scale of $\Delta_T \sim \changed{5} T$ is set by temperature, c.f.\ \eqref{op_g}.
In analogy to \eqref{Sigma_tilde_at_mu_T0}, the flow of $\Sigma(\mu + \Delta\omega)$, with $\Delta \omega \in [-\Delta_T, \Delta_T]$ can be rewritten using  
\begin{widetext}
\changed{
\begin{align}
   \partial_\Lambda (\tilde\Sigma_d)^{R \sigma}_{j i }(\mu + \Delta\omega) &=  
-\frac{2}{\pi} \int^\mu_{-\infty} d\omega' \,  \Imp \Big[ \Big( (a^P)^{\sigma \sigma l k}_{ji}(\mu + \Delta\omega + \omega') - (a^{D*})^{\sigma \sigma l k}_{ji}(\mu + \Delta\omega - \omega') \Big) S^{R \sigma}_{(i+k)(j+l)}(\omega') \Big] \nonumber \\
&\phantom{=\ }  +\frac{1}{\pi} \int d\omega' \, \Big[ \Big\{\coth\Big(\frac{\omega'-\mu + \Delta \omega}{2T}\Big) - [2\theta(\omega'-\mu)-1] \Big\} \Imp (a^P)^{\sigma \sigma l k}_{ji}(\mu + \Delta\omega + \omega') S^{R \sigma *}_{(i+k)(j+l)}(\omega')    \nonumber \\
&\phantom{=\ }  + \Big\{(1-2n_F(\omega')) - [2\theta(\omega'-\mu)-1] \Big\}(a^P)^{\sigma \sigma l k}_{ji}(\mu + \Delta\omega + \omega') \Imp S^{R \sigma}_{(i+k)(j+l)}(\omega')  \nonumber \\
&\phantom{=\ }  - \Big\{\coth\Big(\frac{\mu + \Delta\omega - \omega'}{2T} \Big) - [2\theta(\mu - \omega')-1] \Big\} \Imp (a^{D})^{\sigma \sigma l k}_{ji}(\mu +\Delta \omega - \omega') S^{R \sigma}_{(i+k)(j+l)}(\omega') \nonumber \\
&\phantom{=\ }  - \Big\{(1-2n_F(\omega')) - [2\theta(\omega'-\mu)-1] \Big\} (a^{D})^{\sigma \sigma l k}_{ji}(\mu +\Delta \omega - \omega') \Imp S^{R \sigma}_{(i+k)(j+l)}(\omega') \Big]. 
\label{Sigma_tilde_at_finite_temp}
\end{align}
}
\end{widetext}
Note that in \eqref{Sigma_tilde_at_finite_temp} all four terms in curly brackets $\{\dots\}$  decay exponentially in $\omega'$ on the scale of temperature $T$ for $\omega'$ outside a small interval 
\changed{around $\mu$.} 
%\scrap{with width set again by $T$}.  
Following the same line of argument as above, one finds that the vertex components are %\scrap{exponentially} 
\changed{suppressed }outside of an interval around the feedback frequency which is widened %\scrap{by an amount of} 
\changed{on the} order of %\scrap{the} 
temperature: The important frequencies effectively lie in the intervals
$[2\mu-(\mu-\omega_b)-\tilde\Delta_T,2\mu+\tilde\Delta_T]$ for the P-channel and $[-(\mu-\omega_b)-\tilde\Delta_T,(\mu-\omega_b)+\tilde\Delta_T]$ for the X- and D-channel, where $\tilde\Delta_T \sim \changed{2\Delta_T}$ lies again on the scale of temperature. 
Analogous arguments hold for the complete self-energy.

For finite temperature there is also a two-particle contribution \eqref{tp_g3} to the conductance, directly containing a vertex contribution.    
This vertex contribution is effectively only needed in an interval of width set by temperature around the feedback frequencies. 
This can be seen from \eqref{tp_g3} together with \eqref{definition_phi} and \eqref{def_K} , since the functions 
\begin{align}
f^p(\mu + \Delta_T, \epsilon') = \coth \Big[ \frac{\epsilon' - \mu + \Delta_T }{2T} \Big] - \tanh \Big[ \frac{\epsilon' - \mu}{2T} \Big], \\
f^x(\mu + \Delta_T, \epsilon') = \coth \Big[ \frac{\epsilon' - \mu -\Delta_T}{2T} \Big] - \tanh \Big[ \frac{\epsilon' - \mu}{2T} \Big]
\end{align}
decay exponentially with increasing $|\epsilon' - \mu|$, on a scale set by temperature.
Furthermore, the input argument $\Delta_T$ is analogous to the one appearing in \eqref{Sigma_tilde_at_finite_temp} and lives again on the scale of temperature.
That the leading frequency contribution for the two-particle contribution of the conductance is determined on the scale of temperature can also be nicely seen in Fig.~\ref{fig_dyn_temp_dep_NL}. 
The main contribution to $g_{2}$ is collected by going from $N_L=0$ to $N_L=5$, i.e. while resolving the temperature scale (c.f.\ the discussion in Sec.~\ref{subsection_long_range}). 
Further increase in $N_L>5$ only \changed{slightly} changes the two-particle contribution. % \scrap{slightly}.

\section{Violation of Ward Identities}
\label{vio_ward_identities}
\begin{center}
\begin{figure*}
   \includegraphics[scale=1.0]{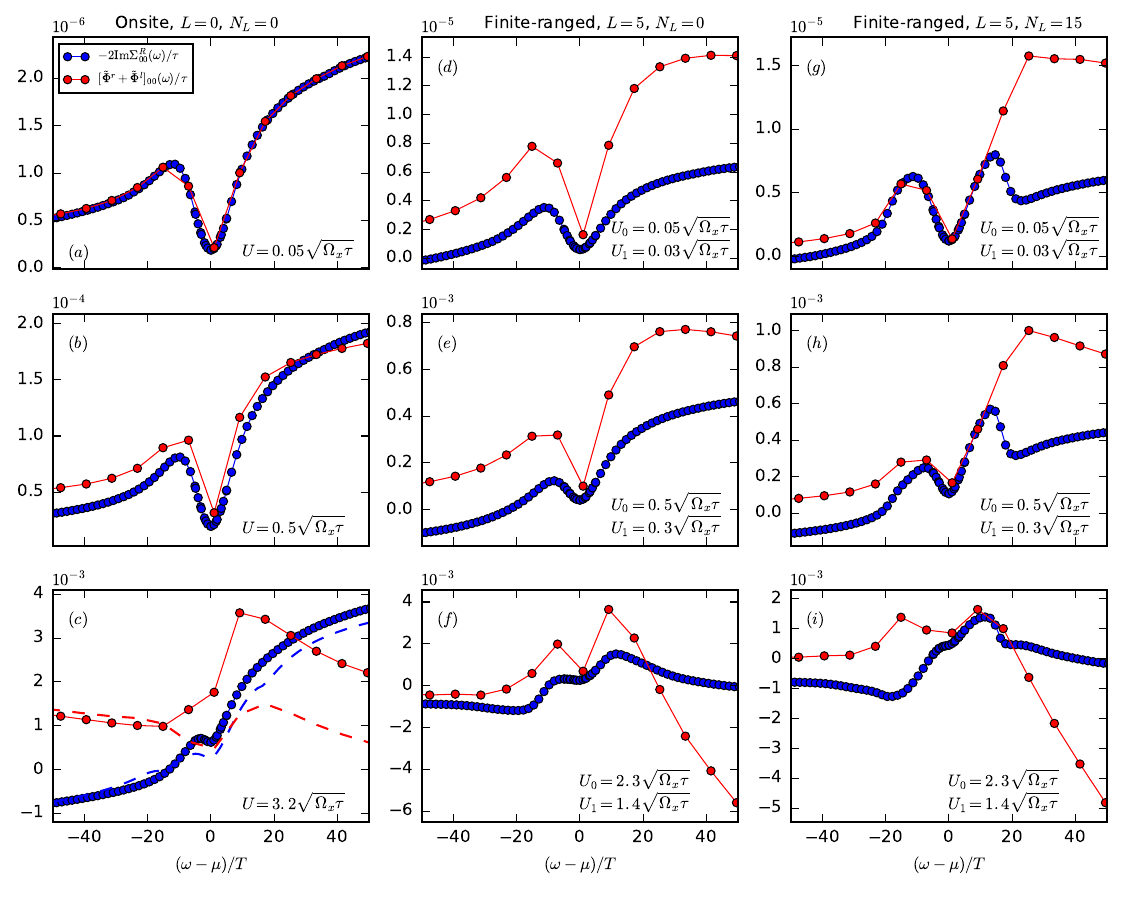} 
   \caption{\small Violation of the Ward identity \eqref{ward_identity} at temperature $T=0.1\Omega_x$ for onsite interactions (first column), and finite-ranged interactions with $N_L=0$ (second column) and $N_L=15$ (third column). The power of $10$ indicated above each panel is a scale factor for the vertical axis. Within each column the interaction strength is increased from very small in the first row, up to the realistic strength in the last row.  
In (c), the dashed lines (blue for $-2\Imp \Sigma^R_{00}$ and red for $(\tilde\Phi^l + \tilde\Phi^r)_{00}$), show the onsite interaction results computed using $L=5$, $N_L=15$. 
Note that with these choices the violation in the region around the chemical potential $\mu$ is reduced compared to the $N_L=0$ result, even in the case of onsite interactions.}
\label{fig_ward_paper}
\end{figure*}
\end{center}

In Sec.~\ref{conductance_computation}, we have seen that the conductance computation suffers from a violation of the Ward identity \eqref{ward_identity}.  
Here, we will elaborate on this violation and show how it depends on external and numerical parameters.
One of the main influences on the severity of this violation are the interaction parameters employed.
For an onsite interaction model our fRG treatment is exact to second order in the interaction, even in the case of the feedback length $L=0$.
Therefore, for small enough interaction strengths, the violation of the Ward identity \eqref{ward_identity} scales like $\sim U^3$, i.e.\ in this weak interaction regime we expect \eqref{ward_identity} to be well satisfied. 
This can indeed be seen in Fig.~\ref{fig_ward_paper}(a,b).

However, for an interaction strength suitable to observe 0.7-physics, the Ward identity is severly violated, see Fig.~\ref{fig_ward_paper}(c). 
For this reason, the best way to obtain the conductance from the results of our current fRG method, is the Ward-corrected treatment described in Sec.~\ref{conductance_computation}, \changed{which restores the} 
%\scrap{By using Eq.~\eqref{ward_symmetrized} we restore} 
Ward consistency between the two-particle part and the self-energy.
%\scrap{ and express as many contributions to the conductance as possible through the self-energy. (We expect the latter to be more accurate than the vertex, since the flow equation for the self-energy involves less approximations than the one of the vertex.)}  

Note that the situation is somewhat remedied by using our eCLA scheme with finite $L$ and finite $N_L$ already for the onsite interaction, see the dashed lines in Fig.~\ref{fig_ward_paper}(c).
In the static Matsubara case \cite{Weidinger2017}, we saw that the eCLA scheme stabilizes the fRG flow by coupling the individual channels better together, extending the accessible physical parameter regime. 
Now we also see that it increases the internal consistency of the results between the one- and two particle level. 

In the case of the model with finite-ranged interactions the situation is qualitatively similar. 
However, with our approximate treatment of the frequency dependence of the long-ranged part of the vertex, described in Sec.~\ref{dynamic_feedback_length}, we generally already make a mistake in second (i.e. the leading order) in the Ward identity. 
This is due to the fact that it is numerically not possible to incorporate the effect of long-range feedback at all frequencies. 
We take long-range contributions only into account in a certain frequency range around the feedback-frequencies [c.f.\ \eqref{static_replacement}]. 
Following the logic of App.~\ref{importance_feedback}, we therefore expect the Ward identity \eqref{ward_identity} to hold only in this frequency range around the chemical potential, even at small interaction strengths. 
This effect can indeed be seen by comparing Figs.~\ref{fig_ward_paper}(d,e) to Figs.~\ref{fig_ward_paper}(g,h).
At large interaction strengths the violation then becomes much more severe, as for the onsite interaction model.
This necessitates introducing the Ward-correction strategy of Eq.~\eqref{ward_symmetrized}.

\changed{
\section{Convergence w.r.t.\ $N_T$} \label{app_Convergence_N_T}
In our whole work, we used $N_T=10$ additional frequencies in the temperature window $[-5T,5T]$ around the chemical potential / feedback frequencies in oder to resolve the finite temperature behavior of the self-energy / two-particle vertex. 
Despite $N_T=10$ being much lesser than the comfortable $\sim 100$  additional frequencies used in Ref.~\cite{Schimmel2017} for the same purpose, our results are still converged w.r.t.\ $N_T$, see Fig.~\ref{fig_NT_dep}. 
Here we compare the results for the finite-ranged interaction model with $N_T=10$ (blue curves) and $N_T=20$ (red curves). 
Note that in order to not change the frequency range $\theta^f_A$, for the respective channels $A \in \{P,X,D \}$, we also had to increase the number of long range frequencies $N_L$ accordingly. 
Both curves lie almost perfectly on top of each other, indicating that a further increase of $N_T$ beyond $10$ is not necessary. 
\begin{figure}
   \includegraphics[scale=1.0]{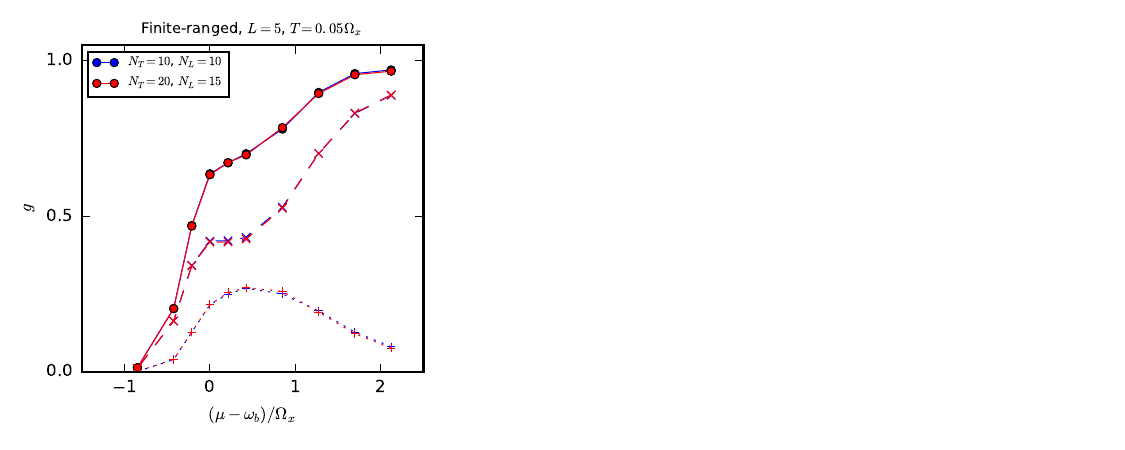} 
   \caption{\small \changed{Conductance curves for $N_T=10$ (blue) and $N_T=20$ (red). As before, solid lines indicate the whole conductance $g$, while dashed lines indicate the one-particle and dotted lines the two-particle contributions. The resulting curves almost perfectly agree.}}
\label{fig_NT_dep}
\end{figure}
}

\bibliography{keldysh_lrfRG_citations}
\end{document}